\documentclass[
aps,%
12pt,%
final,%
notitlepage,%
oneside,%
onecolumn,%
nobibnotes,%
nofootinbib,%
superscriptaddress,%
noshowpacs,%
centertags]%
{revtex4}

\begin{document}

\title{Evolution of a Viscous Protoplanetary Disk
with Convectively Unstable Regions}

\author{\firstname{Ya.~N.}~\surname{Pavlyuchenkov}}
\email{pavyar@inasan.ru}
\affiliation{%
Institute of Astronomy, RAS, Moscow, Russia
}%

\author{\firstname{A.~V.}~\surname{Tutukov}}
\affiliation{%
Institute of Astronomy, RAS, Moscow, Russia
}%

\author{\\ \firstname{L.~A.}~\surname{Maksimova}}
\affiliation{%
Institute of Astronomy, RAS, Moscow, Russia
}%

\author{\firstname{E.~I.}~\surname{Vorobyov}}
\affiliation{%
Southern Federal University, Institute of Physics, Rostov-on-Don, Russia
}%
\affiliation{%
Department of Astrophysics, Vienna University, Vienna, Austria
}%


\begin{abstract}
The role of convection in the gas-dust accretion disk around a young star is studied. 
The evolution of a Keplerian disk is modeled using the Pringle equation, which describes the time variations of the surface density under the action of turbulent viscosity. 
The distributions of the density and temperature in the polar directions are computed simultaneously in the approximation that the disk is hydrostatically stable. The computations of the vertical structure of the disk take into account heating by stellar radiation, interstellar radiation, and viscous heating. The main factor governing evolution of the disk in this model is the dependence of the viscosity coefficient on the radius of the disk. The computations of this coefficient take into account the background viscosity providing the continuous accretion of the gas and the convective viscosity, which depends on the parameters of the convection at a given radius. The results of computations of the global evolution and morphology of the disk obtained in this approach are presented. It is shown that, in the adopted model, the accretion has burst-like character: after the inner part of the disk ( R < 3 AU) is filled with matter, this material is relatively fast discharged onto the star, after which the process is repeated. Our results may be useful for explaining the activity of young FU Ori and EX Lup objects. It is concluded that convection may be one of the mechanisms responsible for the non-steady pattern of accretion in protostellar disks.
\end{abstract}

\maketitle

\section{INTRODUCTION}
Studies of protoplanetary disks of young stars are of
considerable interest. A young star accumulates most
of its mass via the accretion of gas from a circumstellar
disk; i.e., the formation of the star has a direct connection to the evolution of the disk. In turn, a planetary
system is a natural result of the late stages of the evolution of the protostellar disk. An enormous number of
studies have been devoted to the physics and evolution
of protoplanetary disks (see, e.g., the monograph~\cite{2013apf..book.....A}).
Despite great observational and theoretical achievements in the study of protoplanetary disks, many
problems related to the physics of accretion disks have
not been resolved completely. In particular, the question of the mechanism for mass and angular momentum transfer, which enables the accretion of the ambient gas, remains topical. The main approach to
describing disk accretion, along with gravitational
instability, is the formalism of turbulent viscosity, but
the mechanism making the flow turbulent remains under discussion. The convection of accreted gas in
the polar directions may be one of the mechanisms giving rise
to the turbulence.

The idea that convection in the protoplanetary
disks may be responsible not only for heat transfer, but 
also providing viscosity, and, thus, may influence the
evolution of the disk, was formulated in the 1980s~\cite{1978M&P....18....5C,1980MNRAS.191...37L}. 
This idea was greeted with enthusiasm, however,
even after several decades, the role of convection in the
transfer of angular momentum remains under discussion (see the detailed historical review~\cite{2007IAUS..239..405K}). The first
numerical models that took convection into account~\cite{1987Icar...69..423C,1987Icar...69..387C} provided evidence that the viscosity coefficient
corresponding to the convection is very small, and that
the gaseous disk itself tends to break up into rings.
Studies based on later numerical models~\cite{1999ApJ...514..325K,
2003ApJ...582..869K} indicated higher viscosity coefficients (in terms of the
Shakura–Sunyaev parameter $\alpha=10^{-3}-10^{-2}$), however, this required an additional mechanism for heating the gas in the equatorial plane of the accretion
disk. The idea of additional viscosity associated with
convection in circumstellar disks was also discussed in
\cite{2003AN,2014ApJ...787....1H,2016MNRAS.462.3710C}. When high-quality images of protoplanetary
disks were obtained and ring-shaped structures were
observed, interest in hydrodynamical models with
convection again increased. For example, the results
of three-dimensional modeling of convection in a disk
were presented in~\cite{2018MNRAS.480.4797H}. This demonstrated the rise of
convective cells, vortices, and other coherent structures when convection was triggered.
Held and Letter~\cite{2018MNRAS.480.4797H} note that they were unable to obtain a self-sustaining convection regime in the disk. At the same
time, theoretical analyses of the convective instability
of the accretion disk were presented~\cite{2015MNRAS.448.3707S, 
2015MNRAS.451.3995S,2017MNRAS.464..410M}, where
the instability conditions were found and the need for
more detailed models was noted. In the present study,
we have investigated the conditions for the rise of convection and the large-scale evolution of a convective
protostellar Keplerian disk, based on a model featuring a detailed calculation of the vertical structure of
the disk and allowance for the constant accretion of
gas onto the disk from a circumstellar envelope. Our
main focus is the analysis of the recursive nature of the
accretion activity of the protostellar disk.

\section{THE MODEL OF THE DISK}

The evolution of the circumstellar disk was computed using a numerical model in which the evolution
of the radial and vertical structure of the disk is calculated consequentially. Each time step is divided into
two stages: (i) computation of the evolution of the surface density of the gas; (ii) reconstruction of the density and temperature distributions in the vertical direction. In the first stage, the rate of heating of the
medium associated with the accretion of matter,
which is necessary for computing the thermal structure, was also computed. Once the vertical structure of
the disk was reconstructed, convectively unstable
regions were identified and the distribution of the viscosity coefficient needed to compute the evolution of
the surface density was determined. We describe both
stages of our computations in more detail below.

\subsection{Computation of the Evolution of the Surface Density}

To describe the evolution of the surface density of
the disk, we used the formalism of a viscous accretion
disk. This approach assumes that the mechanism for
the transfer of mass and angular momentum is some
physical process (turbulence, magnetic fields, convection), mathematically described in a way analogous to
a molecular viscosity; i.e., the model is based on the
Navier–Stokes equations. In the approximation of an
axially symmetric, geometrically thin Keplerian disk,
neglecting gas pressure gradients in the radial direction, the evolution of the surface density is described
by the Pringle equation~\cite{1981ARA&A..19..137P}:

\begin{equation}
\frac{\partial \Sigma}{\partial t} =
\frac{3}{R}\frac{\partial }{\partial R}\left[\sqrt{R}\frac{\partial}{\partial R}
\left(\nu\sqrt{R}\Sigma\right)\right] + W(R,t),
\label{eq1}
\end{equation}
where $\Sigma$ is the surface density, $R$ the distance to the
star, $t$ time,  $\nu$ the coefficient of kinematic viscosity,
and $W(R,t)$ the rate of matter inflow from the envelope. The Pringle equation is widely used to describe
the long-term evolution of circumstellar disks (see,
e.g., the review~\cite{2011ARA&A..49..195A}). The specifics of its application
depend on the definition of the function $\nu(R)$. In our 
case, $\nu(R)$ is determined phenomenologically and
using the conditions of convective instability.

As a result of the evolution of a viscous disk, thermal energy is released. In essence, the source of the
released thermal energy is the gravitational energy of
the matter; i.e., viscosity may be considered as a mechanism that transforms the gravitational energy of the
accreting gas into heat. The rate of heat release per unit area is~\cite{1981ARA&A..19..137P}:

\begin{equation}
\Gamma_{\rm vis} = \frac{9}{4} \frac{GM_{\odot}}{R^3}\nu \Sigma,
\label{eq_vis}
\end{equation}
where $M_{\odot}$ is the mass of the central star. This heating
was included in the computations of the vertical disk
structure. In essence, it is precisely this heating that is
responsible for the rise of convection.

\subsubsection{Viscosity coefficient in convectively stable
regions}

We assumed the presence of some background mechanism for the transfer of angular momentum in convectively stable regions and specified the
viscosity in the form:
\begin{equation}
\nu_{\rm bg} = \nu_{0}\left(\frac{R}{R_{\rm AU}}\right)^{\beta},
\label{eq_nu}
\end{equation}
where $\beta=1$ and $\nu_{0}=10^{15}$~cm$^2$/s. With this choice of $\beta$,
the surface density distribution for the stationary
solution of the Pringle equation at $R \gg R_{\odot}$,

\begin{equation}
\dot{M}=3\pi\Sigma\cdot \nu_{\rm bg},
\end{equation}
will be inversely proportional to the distance to the
star, i.e., $\Sigma \propto R^{-1}$, in good agreement with observations of protoplanetary disks~\cite{2011ARA&A..49...67W}. For the adopted
value of $\nu_{0}$ and the accretion rate $\dot{M}=10^{-7} M_{\odot}$/yr, the
mass of the disk between 0.1 and 100 AU will be
$M_{\rm disk}=10^{-1} M_{\odot}$. These values also fit the observed
ranges of accretion rates and masses of protoplanetary
disks~\cite{1998ApJ...495..385H}. We will now show that the values $\beta=1$ and
$\nu_{0}=10^{15}$~cm$^2$/s are consistent with the widely used
$\alpha$-parameterization of the turbulent viscosity.
According to Shakura and Sunyaev~\cite{1973A&A....24..337S}, $\nu
= \alpha ñ_{s}H$,
where $c_s$ is the sound speed and $H$ the height of the
disk. Let thermal structure of the disk be completely
determined by heating by the central star. Then,
\begin{equation}
\varepsilon\,\frac{L_{\odot}}{4\pi R^2} ={\sigma} T^4,
\end{equation}
where $L_{\odot}$ is the stellar luminosity, $\varepsilon$ the cosine of the
angle between the normal to the surface of the disk and
the direction to the star, $\sigma$ the Stefan–Boltzmann
constant, and $T$ the effective temperature. The condition of hydrostatic equilibrium of the disk in the vertical direction can be approximated
\begin{equation}
\frac{H}{R} \approx \frac{c_s}{V_{k}},
\label{eq_height}
\end{equation}
where $V_{k}$ is the Keplerian velocity at the radius $R$.
Combining these two equations and using the relation $c_s^2={k_{\rm B}T}/{\mu_{\rm g} m_{\rm H}}$, where $k_{\rm B}$ is Boltzmann's constant, $\mu_{\rm g}$ the mean molecular weight of the matter, and $m_{\rm H}$ the
mass of a hydrogen atom, we obtain:
\begin{equation}
\nu_{\rm bg}=\alpha \nu_{\alpha} \left(\frac{R}{R_{\rm AU}}\right),\,\,\,
\text{where}\,\,\,\,
\nu_{\alpha}=\dfrac{k_{\rm B}R_{\rm AU}}{\mu_{\rm g} m_{\rm H} (GM_{\odot})^{1/2}} 
\left(\dfrac{\varepsilon L_{\odot}}{4\pi \sigma}\right)^{1/4}.
\end{equation}

Assuming $\varepsilon=0.1$ and inserting the values of the
constants, we obtain $\nu_{\alpha}=10^{16}$~cm$^2$/s. Thus, the
parameterization \eqref{eq_nu} we have used for convectively
stable regions corresponds to a viscous-disk model
with $\alpha=0.1$. This high value of $\alpha$ corresponds to the
earliest (less than 1 Myr) stages of the evolution of protoplanetary disks, when the main contribution to the
viscosity is probably made by the self-gravity of the
disk~\cite{2009MNRAS.393.1157C}.

\subsubsection{Coefficient of viscosity in convectively unstable regions}

We will now describe the algorithm we
used to specify the viscosity in convectively unstable
regions. After computing the vertical structure of the
disk using the method described in Section 2.2,
the regions of the disk in which the condition of convective instability was satisfied were identified. An area
was considered to be convectively unstable if the following condition was satisfied~\cite{1959flme.book.....L}:
\begin{equation}
\frac{dT}{dz} < -\frac{g(z)}{c_{P}},
\label{criterion}
\end{equation}
where $g(z)=\dfrac{GM_{\odot}}{R^3}z+4\pi G\sigma(z)$ is the vertical component of the gravitational acceleration at the radial distance~$R$ and height~$z$,
$\sigma(z)=\int\limits_{0}^{z}\rho(z^\prime)\, dz^\prime$ the surface density measured from the equatîr, $c_{P}=\dfrac{7}{2} \dfrac{k_\text{B}}{\mu_\text{g} m_\text{H}}$ the gas specific heat at constant pressure, and $\mu_\text{g}=2.4$ the mean molecular weight. The criterion~\eqref{criterion} corresponds
to restrictions lying at the basis of the model used to
reconstruct the vertical structure. In particular, this
model does not take into account ionization and dissociation of the gas. Further, for every vertical column defined by the radial coordinate $R$, we computed the
mass fraction of the gas $\gamma$ in convectively unstable
regions in this column. The viscosity coefficient determined by convection is
\begin{equation}
\nu_{\rm c} = \gamma\, H V_{\rm c},
\label{eq_nuconv}
\end{equation}
where $H$ is the local disk scale height, determined
using Eq.~\eqref{eq_height}, and $V_{\rm c}$ the characteristic convection
speed. The speed $V_{\rm c}$ was determined in the approximation that all the energy released by the accretion of gas
is transformed into kinetic energy of the gas' convective motion; i.e., the rate of viscous dissipation is equal to the kinetic energy flux:
\begin{equation}
\Gamma_{\rm vis} = \frac{\rho_{0} V_{\rm c}^2}{2} V_{\rm c},
\end{equation}
where $\rho_{0}$ is the equatorial density. The resulting distribution was smoothed in the radial direction using a
Gaussian with width $H$:
\begin{equation}
\tilde{\nu}_{\rm c}(R) =  \frac{\int\limits_{R_{\rm in}}^{R_{\rm out}}
\nu_{\rm c}(r) {\rm e}^{-\frac{(R-r)^2}{2H^2}} dr}
{\int\limits_{R_{\rm in}}^{R_{\rm out}}{\rm e}^{-\frac{(R-r)^2}{2H^2}} dr},
\end{equation}
where $R_{\rm in}$ and $R_{\rm out}$ are the inner and outer boundaries
of the disk. This radial smoothing was carried out to
ensure that the radius of the convective region is no
less than the height of the disk. The local height of the
disk is a natural choice for the smoothing radius, since
$H$ is taken as the characteristic convection length
when defining viscosity in~\eqref{eq_nuconv}.

The total viscosity coefficient was taken as the sum
of the background and convective viscosities:
\begin{equation}
\nu = \nu_{\rm bg} + \tilde{\nu}_{\rm c}.
\label{eq_nuconv_sum}
\end{equation}

Although this approach appears to overestimate the
convective viscosity, we consider it to be an acceptable
initial approximation. In future, we plan to apply mixing-length theory to carry out a more correct estimation of $\nu$ in convectively unstable regions.

\subsubsection{Initial and boundary conditions, model for
accretion from the envelope}

In the model, we use fixed values for the surface density at the inner ($R=0.1$~AU) and outer ($R=100$~AU) boundaries of the disk, $\Sigma=10^{-2}$~g/cm$^2$ and $\Sigma=10^{-5}$~g/cm$^2$, respectively. 
These boundary values of the surface density are three
to four orders of magnitude lower than the values near
these boundaries after the disk settles to a quasi-equilibrium state. We can assume that these boundary
conditions enable the free flow of matter, i.e., that
some effective mechanisms for the extraction of mass
act at the boundaries of the disk. Such conditions certainly strongly affect the behavior of the solution near
the boundary, as follows from the large surface density
gradients near the boundaries, and they also determine the nature of the accumulation of gas in the disk.
The introduction of more complex boundary conditions requires a separate study, which we plan to carry
out in the future.

The evolution of the accretion disk is largely determined by the accretion of the matter from the circumstellar envelope. Simple estimates show that the star
receives most of the matter from the disk, while the
disk itself receives matter from the envelope, the remnant of the molecular cloud. The region of matter
accretion from the envelope to the disk depends on the
initial angular momentum of the cloud. To estimate the "centrifugal" radius $R_{\text{acc}}$, at which the matter from
the envelope accretes, we used the formula from~\cite{2006ApJ...645L..69D}
\begin{equation}
R_{\text{acc}} = \frac{\Omega^2 R^4_{\text{core}}}{GM_{\text{core}}},
\label{r_acc}
\end{equation}
where $\Omega$ is the angular velocity of the cloud, $R_{\text{core}}$ the
initial position of the accreting element in the cloud,
and $M_{\text{core}}$ the mass of the inner part of the cloud (the
mass of the star). This formula is obtained from the
condition of conservation of angular momentum of
the accreting matter. The density distribution in
observed protostellar clouds is usually described by the
expression
\begin{equation}
n=\dfrac{n_0}{1+\left(\dfrac{r}{r_0}\right)^2},
\end{equation}
where $n_0$ is the central number density of hydrogen and $r_0$ the radius of the inner region with almost constant
density. One example of a well studied protostellar
cloud is the prestellar core L1544, with $r_0 = 3\times 10^3$~AU and $n_0 = 1.6 \times 10^6$~cm$^{-3}$~\cite{2019A&A...623A.118C}. Taking an accreting element located at the boundary of the plateau ($R_{\text{core}}=r_0$), which in L1544 contains a considerable
fraction of the cloud ($M_{\text{core}} = 1.2M_{\odot}$), and assuming
an angular velocity of~$8.23\times10^{-14}$~s$^{-1}$, corresponding
to data for L1544~\cite{2014ApJ...780..188K},
the centrifugal radius $R_{\text{acc}}$ will be 11.3~AU.
Obviously, more distant elements of the
cloud should accrete at larger centrifugal radii. In our
model, the accretion flow from the envelope onto the
disk $W(R,t)$ is specified approximately, namely, we
assumed a constant influx of matter into a ring
between 10 and 20~AU at the rate $10^{-7} M_{\odot}$/yr. We
chose a constant matter-influx rate because we considered only the initial stages of the disk evolution. At the initial time, we specified an initial distribution of the surface density $\Sigma(R) \propto R^{-1}$ and mass $10^{-7}M_{\odot}$ of a disk around a solar-mass star. This initial disk has only
a nominal, formal nature, since the mass of the disk
after reaching a quasi-equilibrium state will significantly exceed this value. In fact, in our formulation of
the problem, the protoplanetary disk is formed completely via the accretion of gas from the envelope.

\subsubsection{Method used to solve the Pringle equation}

We solved the Pringle equation using a finite-difference method. The computational domain was split
into cells, and Eq.~(\ref{eq1}) was approximated in a fully
implicit form:
\begin{eqnarray}
\frac{\Sigma_{i}^{(n+1)}-\Sigma_{i}^{n}}{\Delta t} = &&
\frac{3r_{i+1}^{1/2}}{R_{i}\Delta r_{i} \Delta R_{i+1}}\left(
\nu_{i+1}R_{i+1}^{1/2}\Sigma_{i+1}^{(n+1)}-\nu_{i}R_{i}^{1/2}\Sigma_{i}^{(n+1)}\right)\nonumber \\
&&-\frac{3r_{i}^{1/2}}{R_{i}\Delta r_{i} \Delta R_{i}}\left(
\nu_{i}R_{i}^{1/2}\Sigma_{i}^{(n+1)}-\nu_{i-1}R_{i-1}^{1/2}\Sigma_{i-1}^{(n+1)}\right),
\label{eq2}
\end{eqnarray}
where $\Sigma_{i}^{(n)}$ is the surface density in cell $i$ for time step $n$,
$\Delta t$ the time step, $r_{i}$ the left-hand border of cell $i$,
$R_{i}$ the center of cell $i$, $\Delta r_{i}$ the linear scale of cell $i$, and $\Delta R_{i}$ the distance between the centers of cells $i$ and ($i-1$):
\begin{eqnarray*}
&&R_{i}=\frac{1}{2}\left(r_{i}+r_{i+1}\right)\\
&&\Delta r_{i}=\left(r_{i+1}-r_{i}\right)\\
&&\Delta R_{i}=\left(R_{i}-R_{i-1}\right).
\end{eqnarray*}
Equations~(\ref{eq2}) form a system of linear algebraic equations (with respect to the unknowns $\Sigma_{i}^{(n+1)}$) of the form:
\begin{equation}
A_{i}\Sigma_{i-1}^{(n+1)}+B_{i}\Sigma_{i}^{(n+1)}+C_{i}\Sigma_{i+1}^{(n+1)}=D_{i}.
\end{equation}
Together with the equations realizing the boundary
conditions, these equations form a system of linear
algebraic equations with a tridiagonal matrix. The
solution of this system was found using the tridiagonal
matrix method. The use of an implicit scheme to
approximate the initial equation enabled us to make a
reasonable choice of the time step based on the
adopted accuracy, since this method is absolutely stable. In our computations, the time step was selected
based on the rate of variation of the surface density. In
particular, the time step was decreased when convectively unstable regions were identified.


\subsection{Vertical Structure of the Disk}
\label{sect_vert}

To compute the vertical structure of the disk, we
used the thermal model of~\cite{2017A&A...606A...5V}. 
In this model, the
one-dimensional radiative-transfer problem in the
vertical (polar) direction is solved, taking into account
heating by internal sources and external radiation, as
well as the transfer of thermal radiation in the disk
itself. The model assumes that the only source of
opacity is dust, and that the temperatures of the gas
and dust are equal. The dust-to-gas ratio was assumed
to be constant throughout the disk and equal to $10^{-2}$.
To solve this problem, the radiation was separated into
ultraviolet (stellar and interstellar) and infrared (thermal) radiation of the disk itself. The intensity of the
UV radiation was found via direct integration of the
radiative-transfer equation. The heating of the disk by the UV radiation of the star can be found from the
formula:
\begin{equation}
S_{\rm star} = \dfrac{\kappa_{\rm P}^{\rm uv}L}{4\pi R^2} \exp \left(-\tau_{\rm uv}/\cos\theta\right),
\end{equation}
where $\kappa_{\rm P}^{\rm uv}$ is the Planck-averaged absorption coefficient for the temperature of the star, $L$ the stellar luminosity, $\tau_{\rm
uv}$ the optical depth to UV radiation in the vertical direction from a given position to the upper
boundary of the disk, and $\cos\theta$ cosine of the angle
between the normal and the direction toward the star.
The luminosity of the star was taken equal to be the
sum of the photospheric and accretion luminosities.
The luminosity of the photosphere was assumed to be
equal to the solar luminosity, and the accretion luminosity was calculated from the rate of accretion of
matter from the disk onto the star. When computing
$\kappa_{\rm P}^{\rm uv}$ and $\tau_{\rm uv}$, we used the stellar radiation temperature $T_{\rm star}=6000$~K. The quantity $\cos\theta$ was specified to be
constant over the disk and equal to 0.05. This approximation substantially simplifies the computation of
the external heating, but it does not describe the
effects of self-shadowing that can arise if the density is
distributed non-monotonically. The heating by interstellar UV radiation was calculated using the formula:
\begin{equation}
S_{\rm bg} = D \kappa_{\rm P}^{\rm uv} \sigma T_{\rm bg}^4 \exp \left(-2\tau_{\rm uv}\right),
\end{equation}
where $T_{\rm bg}=10^4$~K and $D=10^{-14}$ are the temperature
and dilution of the interstellar radiation and $\sigma$ is the
Stefan–Boltzmann constant. Note that the contribution of heating by interstellar radiation is insignificant
for the regions under consideration, compared to
heating by the central star, and can therefore be
neglected.

To simulate the transfer of thermal radiation, a system of moment equations was solved in the Eddington
approximation, using frequency-averaged absorption
coefficients:
\begin{eqnarray}
\label{m1}
&&c_{\rm V} \dfrac{\partial T}{\partial t} 
= \kappa_{\rm P} c (E-aT^4) + S \label{m1} \\
&&\dfrac{\partial E}{\partial t} - \dfrac{\partial}{\partial z}
\left(\dfrac{c}{3\rho\kappa_{\rm R}} \dfrac{\partial E}{\partial z}\right) 
= -\rho \kappa_{\rm P} c(E-aT^4),
\label{m2}
\end{eqnarray}
where $T$ is the temperature of the medium, $E$ the radiative energy density, $z$ the vertical coordinate, $\rho$ the
volumetric density, $c_{V}$ the specific heat of the
medium, $c$ the speed of light, $a$ Stefan's constant,
$\kappa_{\rm P}$ the Planck-mean opacity,
$\kappa_{\rm R}$ the Rosseland-mean opacity, and
$S$ the heating function (per unit mass) for
heating by stellar and interstellar radiation, as well as
viscous friction:
\begin{equation}
S=S_{\rm star} + S_{\rm bg} + \frac{\Gamma_{\rm vis}}{\Sigma}.
\end{equation}

Despite the fact that the initial thermal model is
non-stationary, the solution of the system \eqref{m1}--\eqref{m2}
was found in a steady-state approximation. In other
words, it was assumed that the time to reach thermal
equilibrium in the vertical direction is significantly
lower than the timescale for the viscous evolution at
the given radius. Our test computations and estimates,
similar to those presented in~\cite{2014ARep...58..522V}, showed that this
approximation works well for the adopted model.
However, for other parameters of the model, the
steady-state condition could be violated in convectively unstable regions, where the characteristic time
to reach thermal equilibrium and the viscous-evolution timescale can be comparable. Therefore, in the
general case, a non-stationary model should be used.
However, the use of a non-stationary thermal model
will also require the computation of the convective
heat transfer in the radial direction, which is beyond
the scope of the 1 + 1D approach used here.

\begin{figure}
\setcaptionmargin{5mm}
\onelinecaptionstrue
\includegraphics[width=0.49\textwidth]{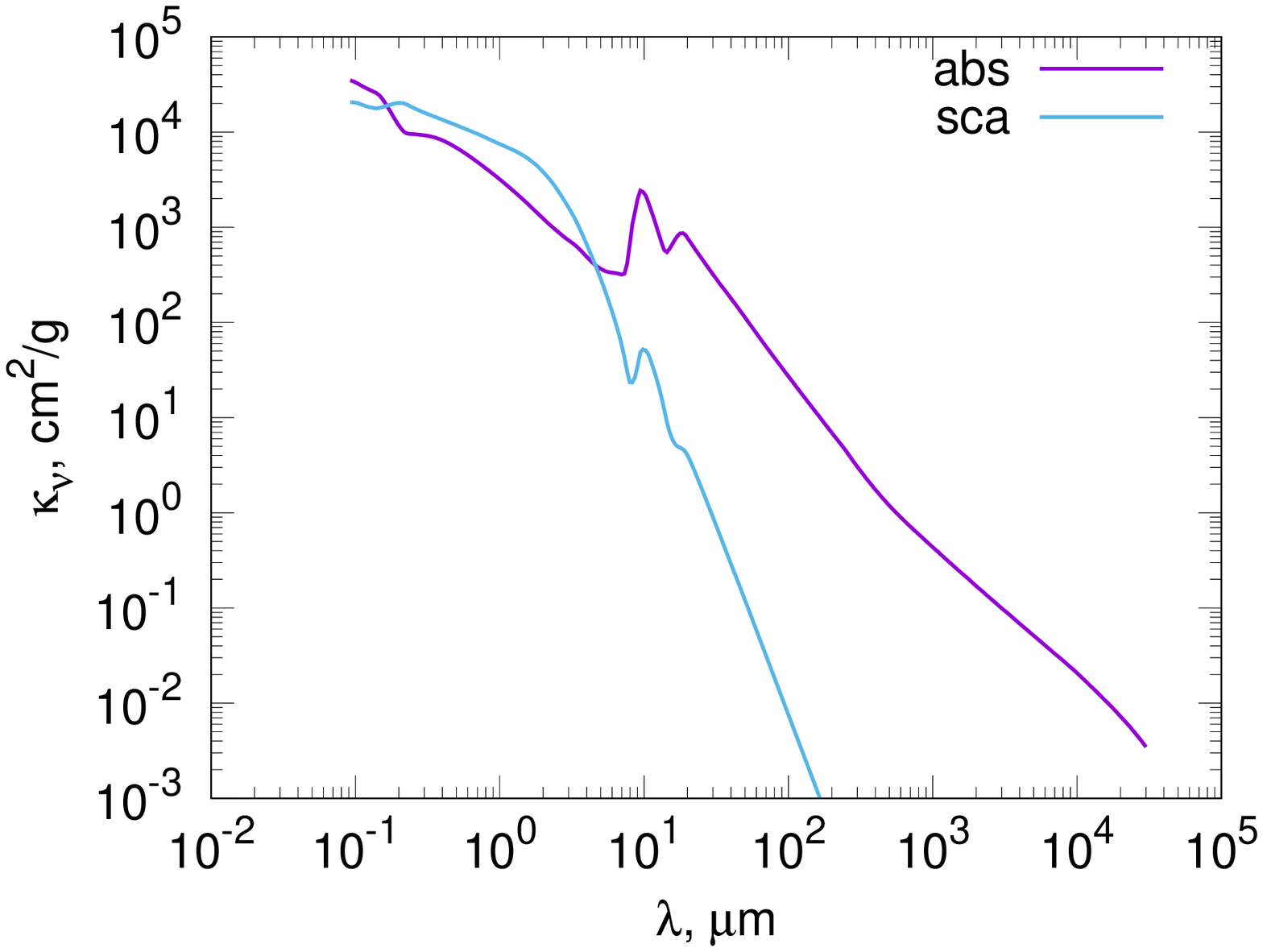}
\hfill
\includegraphics[width=0.49\textwidth]{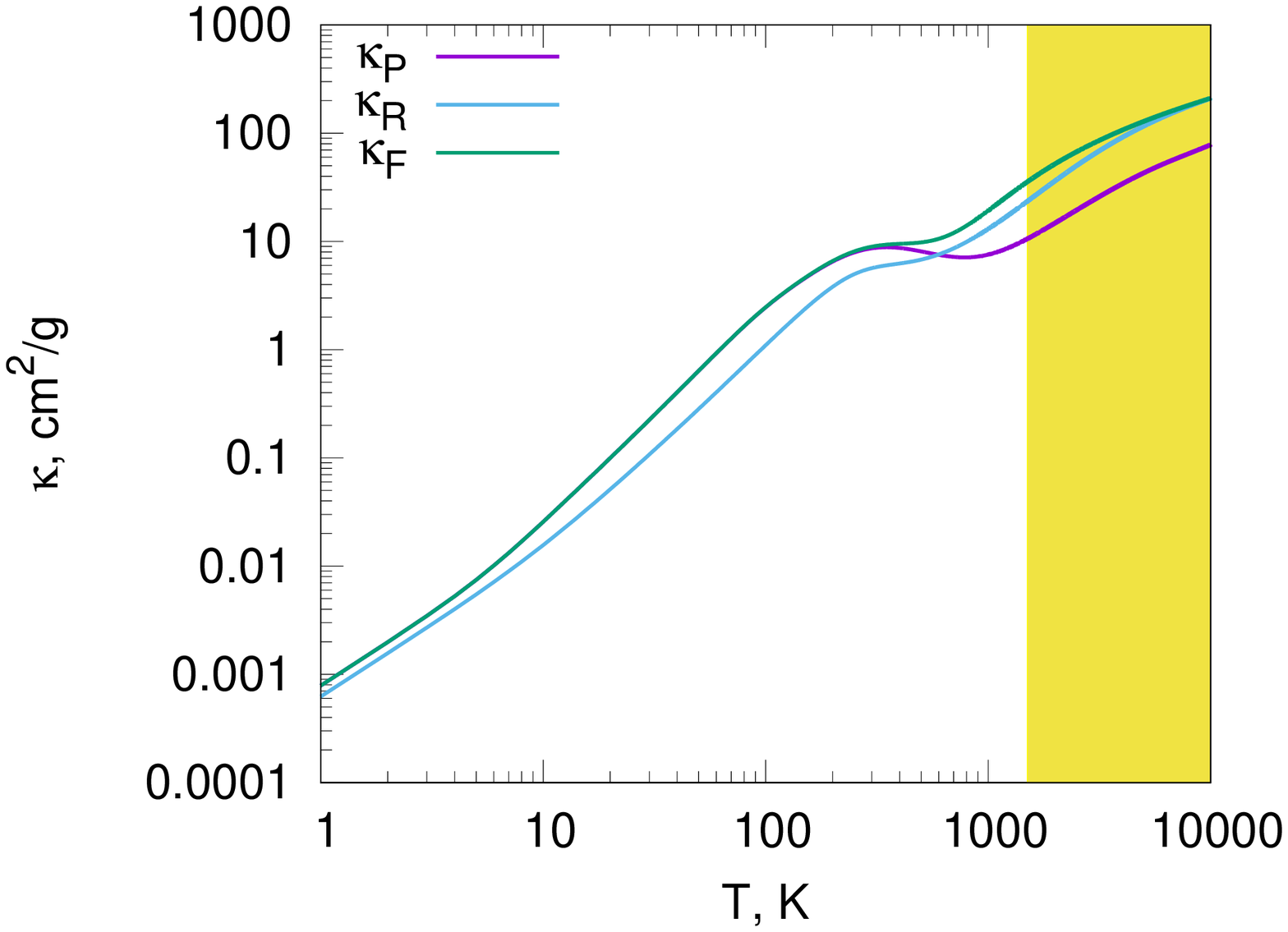}
\captionstyle{normal}
\caption{Wavelength dependence of absorption and scattering coefficients for a mixture of silicate and graphite dust particles (left
panel) and the corresponding Plank-mean $\kappa_{\rm P}$, Rosseland-mean $\kappa_{\rm R}$ and flux-mean $\kappa_{\rm F}$ absorption coefficients (right panel).}
\label{figopac}
\end{figure}

An important feature of the thermal model is the
use of the temperature-dependent Planck-mean and
Rosseland-mean opacities. These opacities were computed from the frequency-dependent absorption and
scattering coefficients for a mixture of graphite and silicate dust particles. The spectral absorption and scattering coefficients themselves were computed using
Mie theory, taking size distribution of dust particles to
be a power law, $n(a)\propto a^{-3.5}$, with the minimum and
maximum radii of the dust particles $a_{\rm min}=5\times 10^{-7}$~cm
and $a_{\rm max}=10^{-4}$~cm. The wavelength-dependent and corresponding Planck-mean, Rosseland-mean, and external flux-averaged opacities are shown in Fig.~\ref{figopac}. An
important feature of the mean opacities is their growth
with temperature at $T>500$~K. As our computations
show, this is one of the conditions for the appearance
of convectively unstable regions.

Note, our models do not account for evaporation
of the dust at $T \gtrsim 1500$~K, when there should be a
sharp decrease in opacity. At such high temperatures, the gas starts to provide the main input to
the opacity. Moreover, at $T \gtrsim  2000$~K, dissociation and
ionization of the hydrogen become substantial, but
they are likewise not included in the model. In Fig.~\ref{figopac},
the temperature range where the opacities we have
used are, strictly speaking, incorrect, is marked in yellow. In our models, the temperature of the medium is
usually below this critical range, but it can approach
these values at the maximum temperature, leading to
the need to modify the model. Such modifications will
significantly complicate the thermal model, and we
plan to carry out such a modification in the future.

The solution of the system of equations~\eqref{m1}--\eqref{m2}
was determined using a fully implicit scheme, which
makes the method absolutely stable and removes
restrictions on the time step. This method enabled us
to correctly compute the thermal evolution in all
regions of the disk, including optically thick regions, where the characteristic heating and cooling timescales are comparable to the dynamic timescales. The
method used to model the thermal structure is closely
related to the method used to reconstruct the vertical
structure of the disk assuming local hydrostatic equilibrium, which was found from the equation:
\begin{equation}
\frac{k_{B}}{\mu_{\rm g} m_{\rm H}}
\dfrac{d(\rho T)}{\rho\, dz} =  -\dfrac{G M_\ast}{r^3}z - 4\pi G \sigma,
\label{static}
\end{equation}
where $M_\ast$ is the stellar mass and $\sigma$ the surface density 
measured in the direction from the equator. The first
term on the right-hand side of~\eqref{static} takes into account
the vertical component of the stellar gravitational
field, and the second term takes into account the self-gravity of the disk. Note that, with the model parameters used, the self-gravity of the disk can be neglected.
The computation of the vertical structure of the disk
enabled us to obtain complete information about the
distribution of the density and temperature in the disk.
We also used a stable implicit method to solve the
equation of hydrostatic equilibrium equation.

The fundamental condition for the efficiency of
these methods is optimal choice of the spatial grid in
the $z$ direction. The spatial grid must track all previously unknown features of the solution (the density
and temperature gradients), taking into account the
existence of a significant restriction on the number of
cells (no more than 100) and the large intervals of the
gas density (up to 10 orders of magnitude). We have
developed an algorithm for the construction and
adaptive modification of such a grid, based on an
approximate fast solution of the equation of hydrostatic equilibrium. This method for the reconstruction
of the vertical structure of the disk taking into account
radiative transfer was thoroughly tested and compared
with other methods. In the stationary regime, the temperature distributions are in good agreement with the
results of modeling of the disk structure obtained in other studies. In the 
non-stationary regime, the characteristic times for establishing thermal equilibrium
correspond to analytical estimates. A more detailed
description of this method can be found in~\cite{2017A&A...606A...5V}.

\section{RESULTS OF THE MODELING}

\begin{figure}
\setcaptionmargin{5mm}
\onelinecaptionstrue
\includegraphics[angle=0,width=0.48\textwidth]{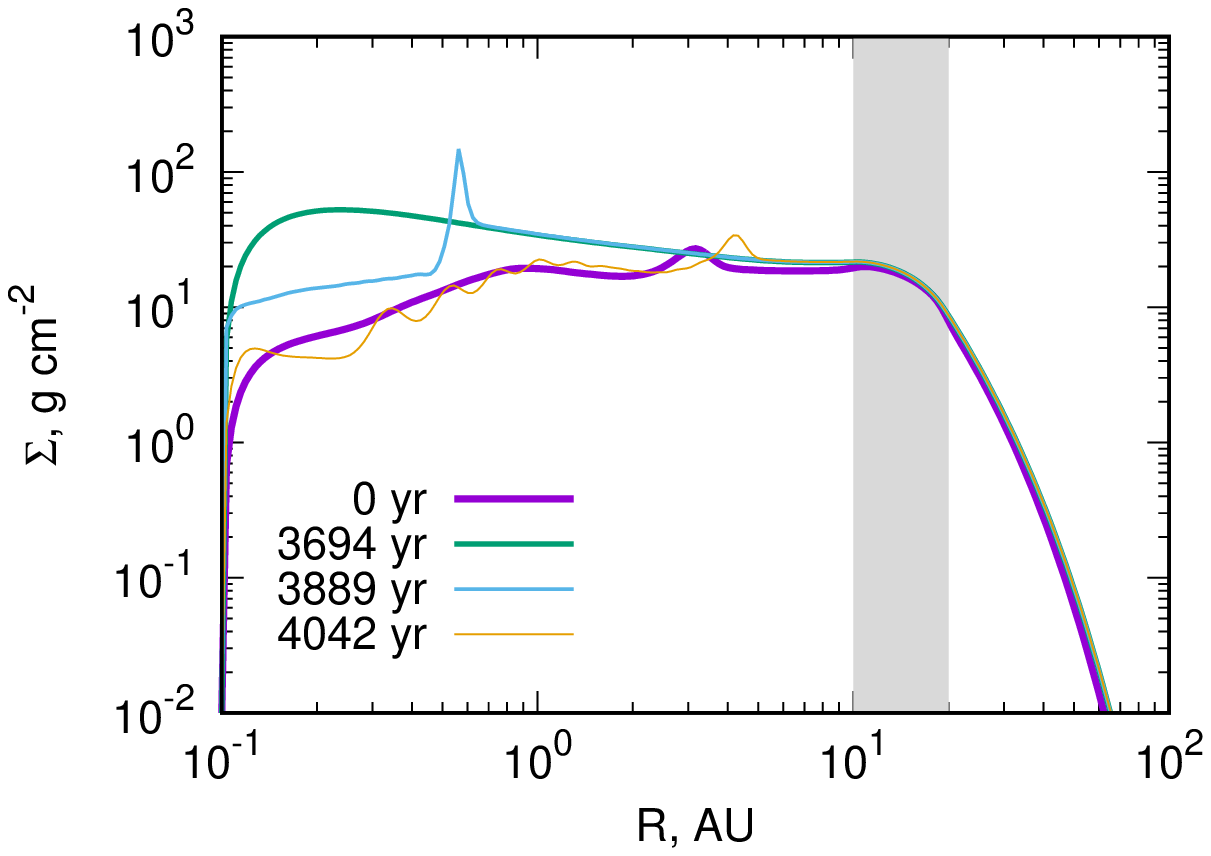}
\includegraphics[angle=0,width=0.48\textwidth]{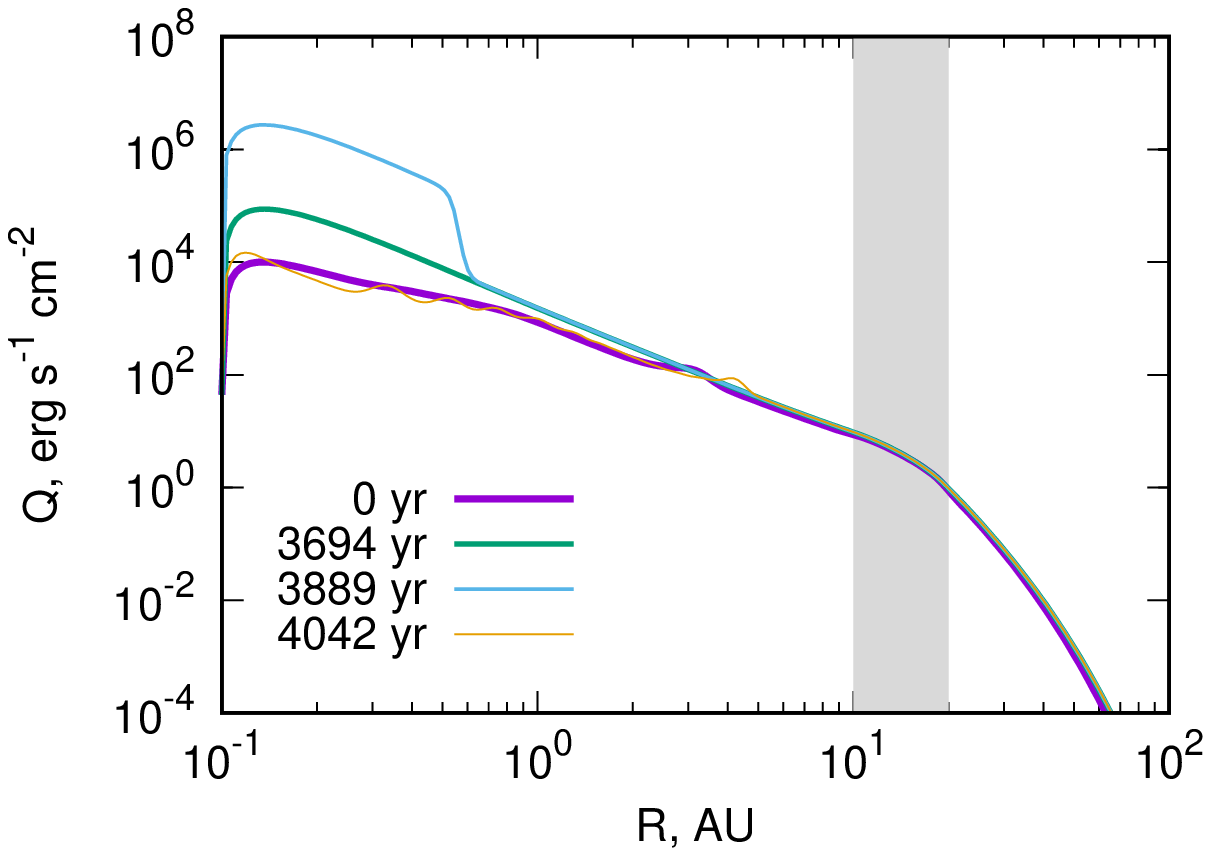}\\
\includegraphics[angle=0,width=0.48\textwidth]{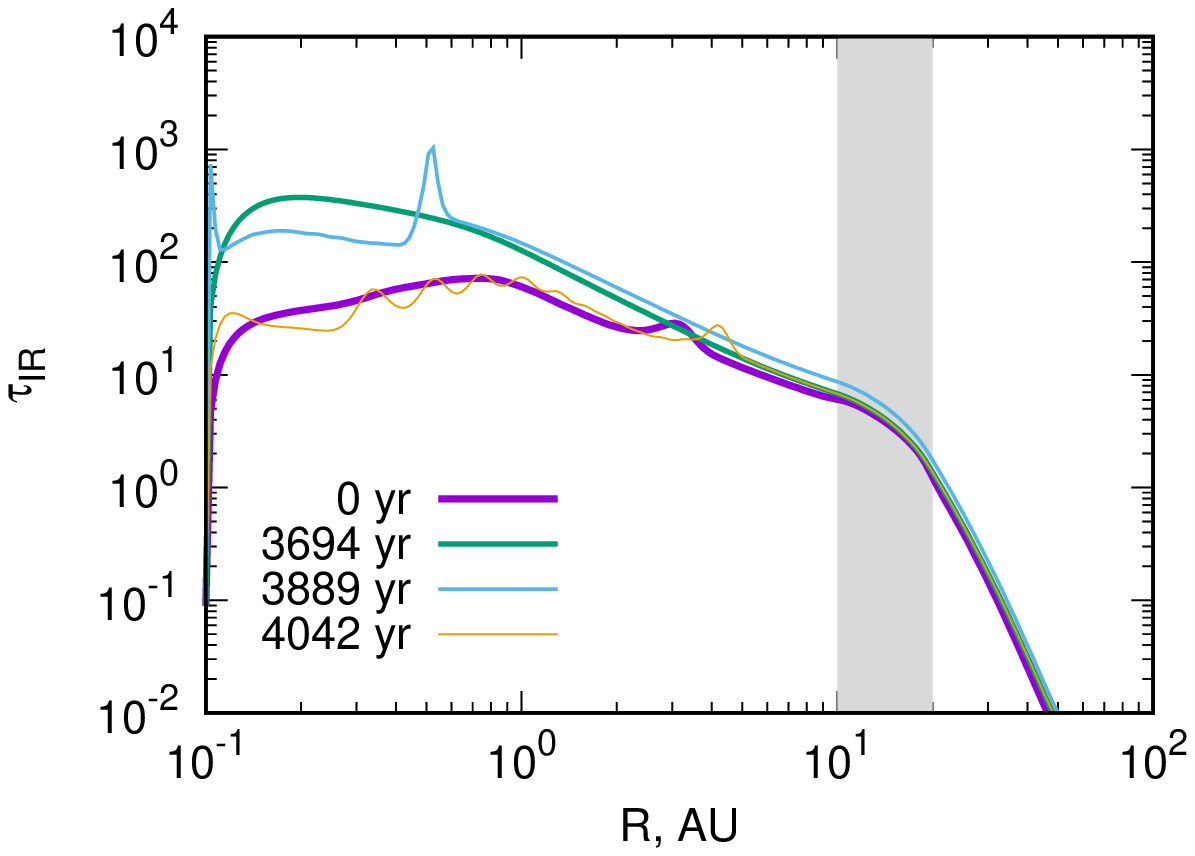}
\includegraphics[angle=0,width=0.48\textwidth]{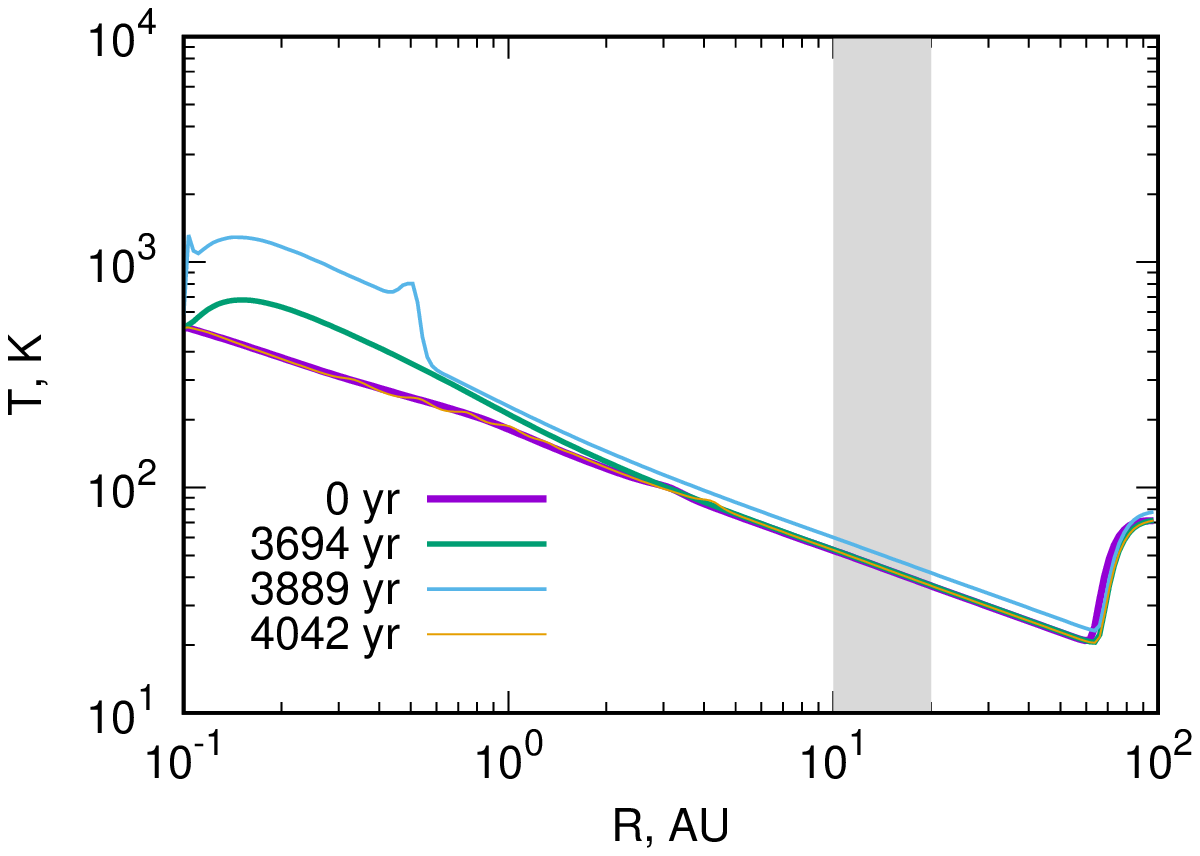}\\
\captionstyle{normal}
\caption{
Radial distributions of the surface density (upper left panel), viscous heating rate (upper right panel), optical depth in the
infrared (lower left panel), and equatorial temperature (lower right panel) for several times, illustrating the development of an
accretion burst. Time is measured from the end of the previous accretion burst. The vertical strip shows the region of gas accretion
from the envelope.}
\label{fig_1d}
\end{figure}

We will now consider the simulation results for
35000 yr after the onset of the evolution of the adopted
disk model. Starting from this time, a periodic, burst-like character of the accretion is established: the inner
region of the disk ($R <$ 3 AU) is gradually filled with
matter, followed by a relatively rapid transfer of matter
from the inner region of the disk to the star due to convection, after which the process is repeated. Further,
this time will be arbitrarily taken to be the zero point in
time. Fig.~\ref{fig_1d} shows the evolution of the distributions
of the gas surface density, the rate of viscous heating,
the optical depth to infrared radiation, and the temperature during an ordinary cycle of this burst accretion regime.

At the initial (zero) time, the surface density in the
inner region $R <$ 3 AU increases in the radial direction
from the star as a consequence of the transfer of matter
after the previous cycle. The equatorial temperature of
the disk monotonically decreases from the inner edge
of the disk to $R\approx 60$~AU, then experiences a jump.
This jump at the outer boundary is due to the fact that
the surface density is so small that the disk becomes
transparent to the UV radiation of the star in the vertical direction; i.e., in the one-dimensional approximation applied, the equatorial regions are heated by the stellar radiation directly.

The inner region ($R <$~3~AU) is gradually filled
with the accreted matter. At $t = 3694$~yr, the density 
distribution in the inner region becomes more monotonic. The temperature in the inner region increases.
This is associated with an increase in the surface density and an accretion flow in the region. The optical depth in the infrared increases to  $\tau_\text{IR}\approx 400$ at the radius $R=0.2$~AU.

At this point, the inner region becomes convectively unstable, the region of convective instability
propagates outward, and a density maximum forms at
its front. At time $t = 3889$~yr, the front reaches the
radial distance 0.5~AU. The energy release $Q$ inside a
radius of 0.5~AU at $t = 3889$~yr is approximately two
orders of magnitude higher than it was before the formation of the convectively unstable zone (Fig.~\ref{fig_1d}). The
temperature inside the convection zone at this time
increases substantially, reaching 1000~K at a radius of
0.2~AU. Beyond the boundary of the convective
region, the temperature also rises compared to the
temperature characteristic for the previous time
period. This is a consequence of the addition to the
stellar photospheric luminosity of significant heat
energy due to the enhanced accretion of the disk material by the star. At $t = 4042$~yr, the expanding convective front attains the radius $R = 3$~AU. The surface
density inside this radius decreases significantly, compared to the time before the outburst, while the density
distribution itself becomes close to the initial distribution for time zero, with the difference that weak oscillations are visible in the distribution, which are later
smoothed out. As Fig.~\ref{fig_1d} shows, at $t = 4042$~yr, the
other distributions are also close to the corresponding
distributions at time zero. From this moment on, the
inner region becomes stable to convection and the disk
enters a new stage of matter accumulation.

\begin{figure}
\setcaptionmargin{5mm}
\onelinecaptionstrue
\includegraphics[angle=270,width=0.48\textwidth]{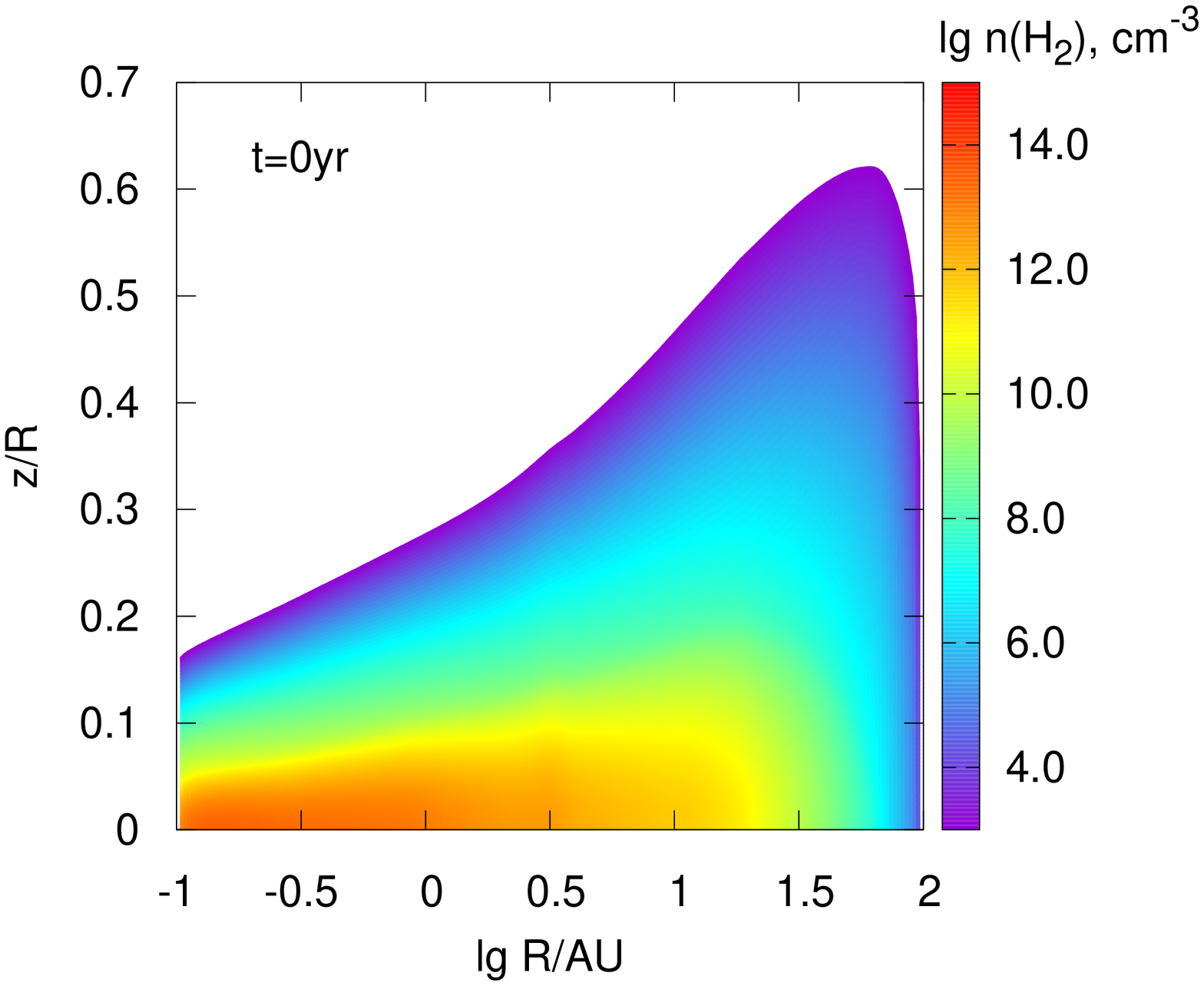}
\includegraphics[angle=270,width=0.48\textwidth]{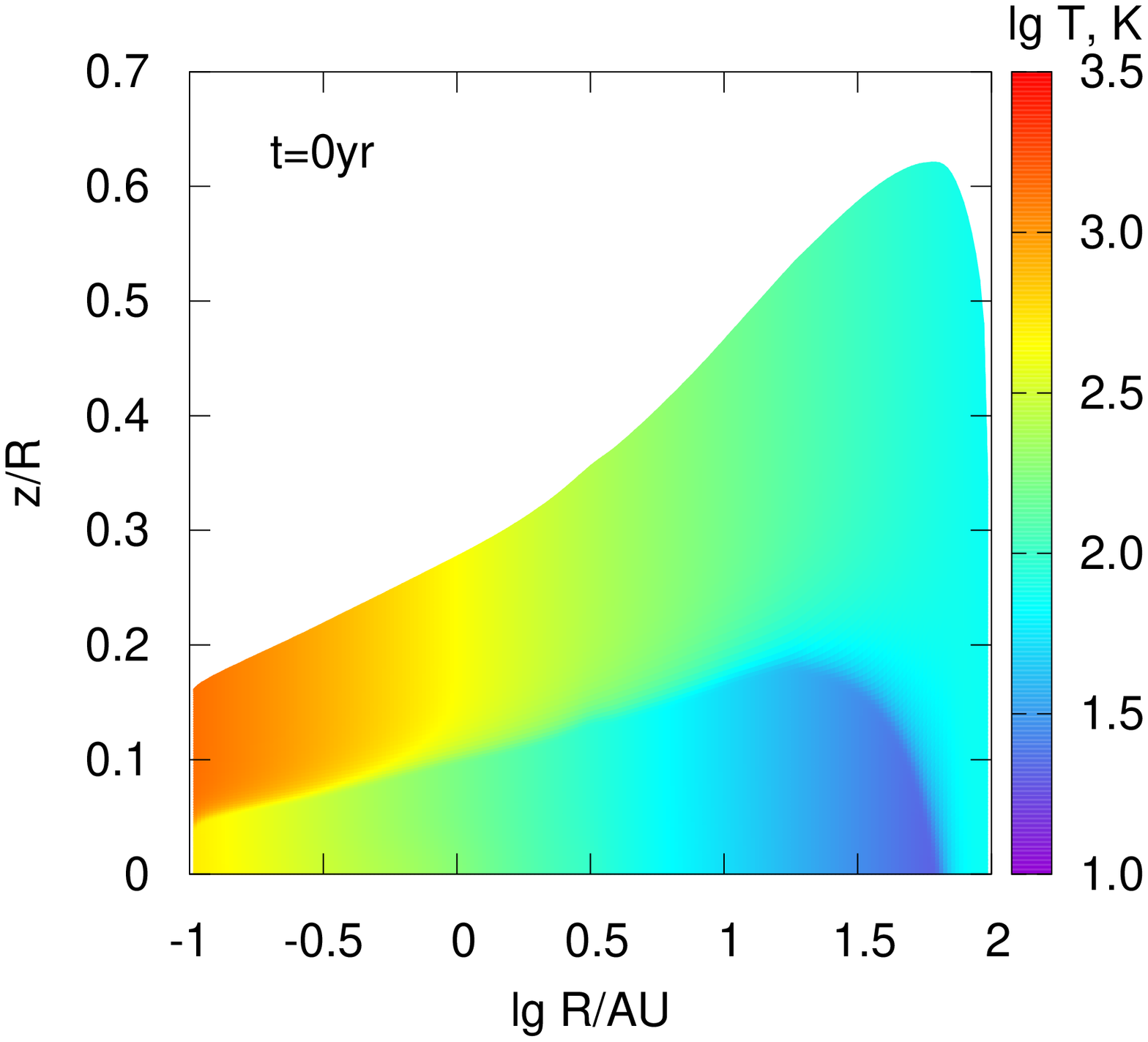}\\
\includegraphics[angle=270,width=0.48\textwidth]{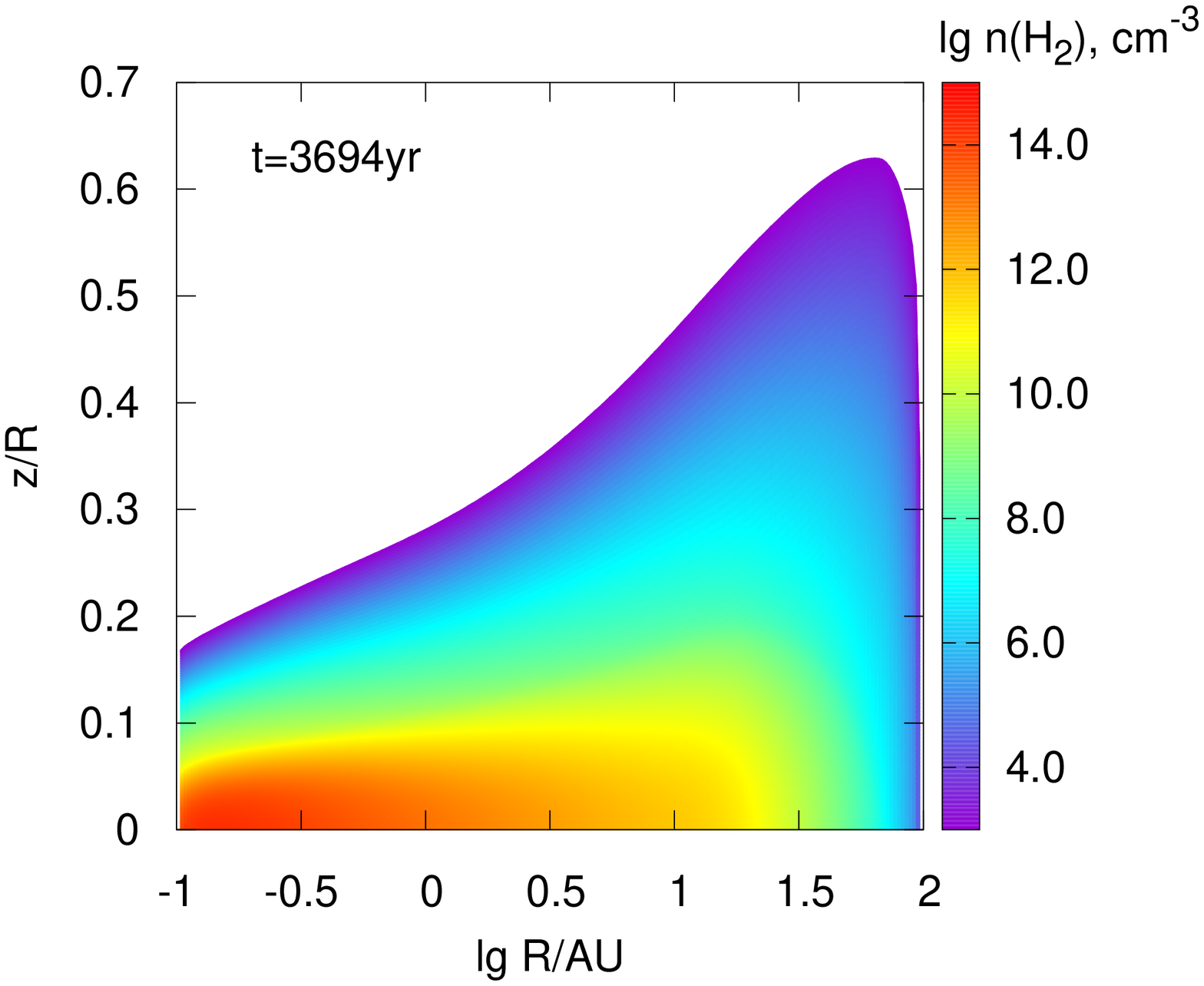}
\includegraphics[angle=270,width=0.48\textwidth]{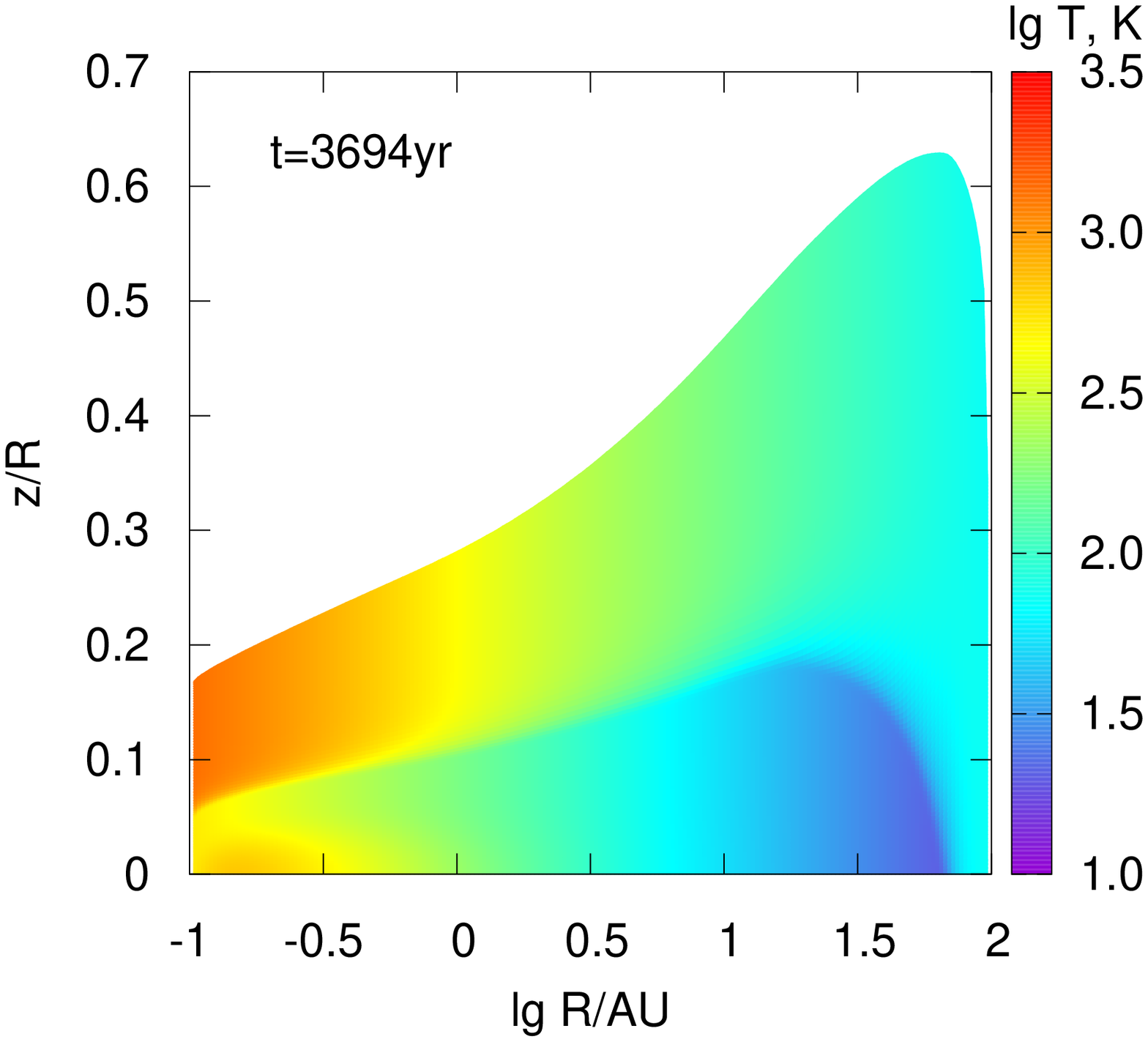}\\
\includegraphics[angle=270,width=0.48\textwidth]{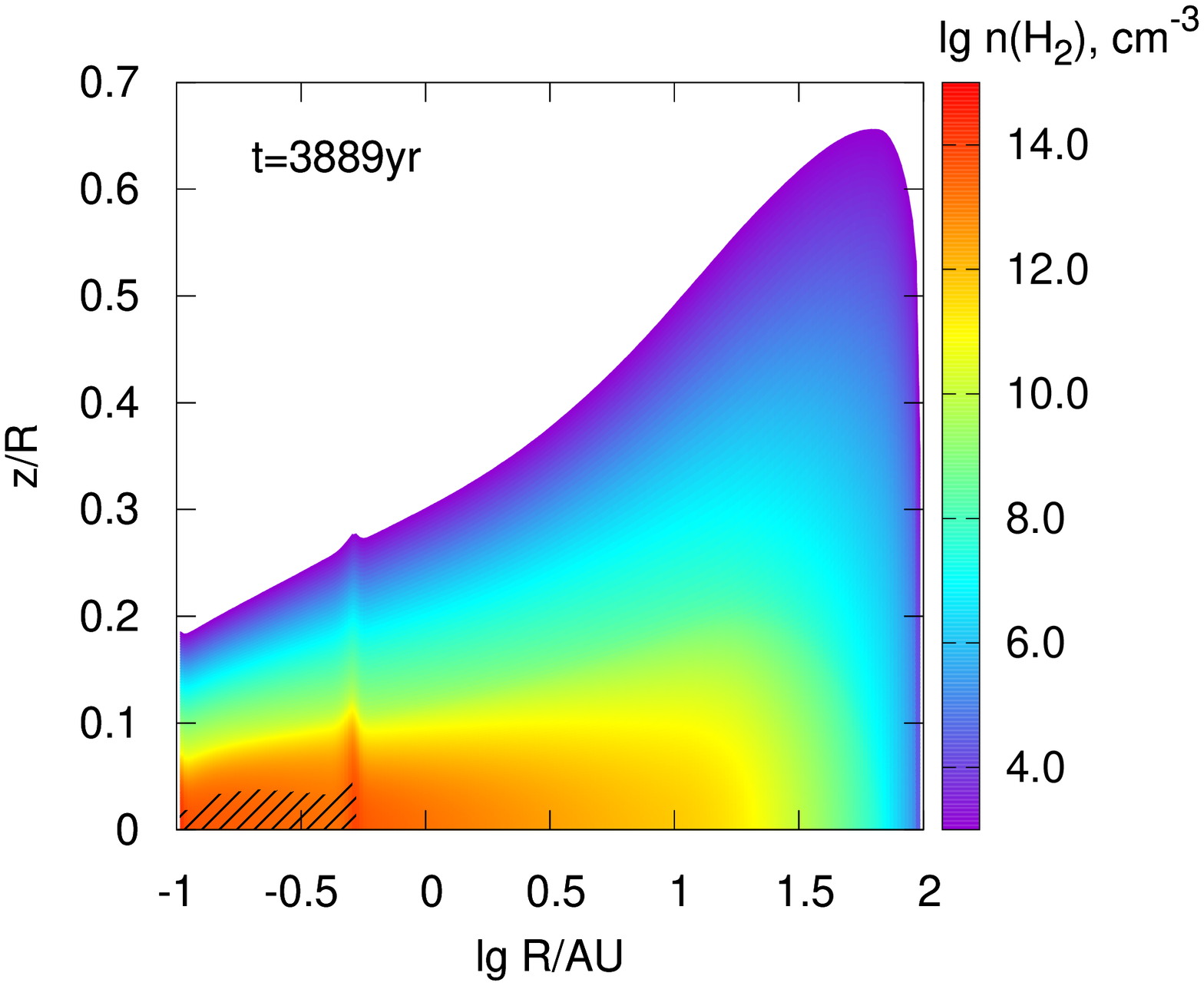}
\includegraphics[angle=270,width=0.48\textwidth]{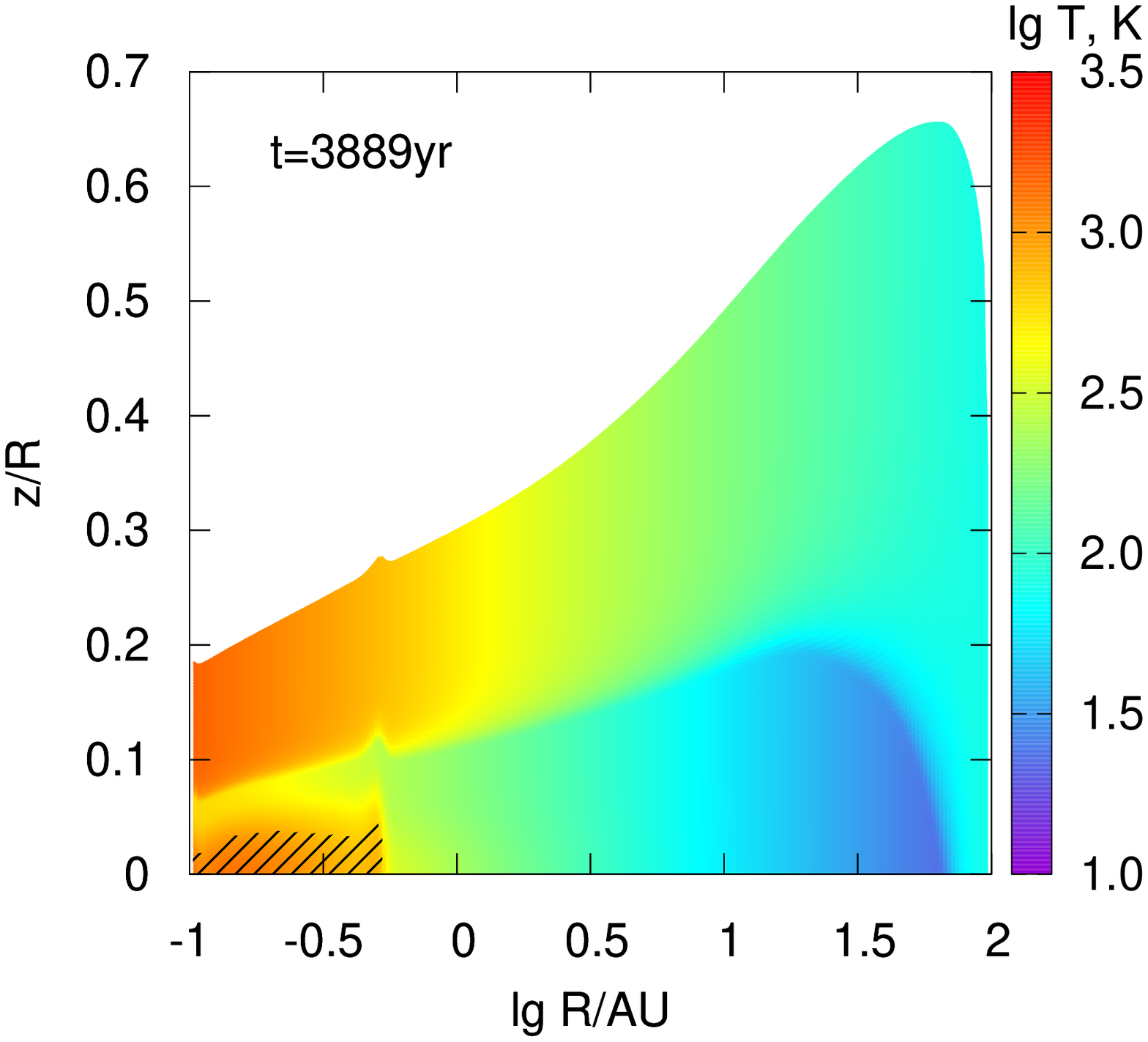}\\
\captionstyle{normal}
\caption{
Distributions of the gas number density (left panels) and gas temperature (right panels) in an RZ cross section of the disk
for three times illustrating the development of an accretion burst. The time is measured from the end of the previous accretion
burst and is indicated in the upper left corner of each panel. The vertical axis shows the ratio of the height above the equator to the
radial distance to the star. Convectively unstable regions are shaded.}
\label{fig_2d}
\end{figure}

Figure~\ref{fig_2d} shows the density and temperature distributions in the polar cross sections of the disk for three
times, illustrating the development of an accretion
outburst.
The panels in the upper row of Fig.~\ref{fig_2d} show
the distributions for the zero time (the beginning of the
cycle). The gas density in the entire disk decreases
monotonically from the equator to the upper boundary of the disk. Near the inner boundary of the disk,
the density contrast of the disk in the polar direction
reaches $\sim$11 orders of magnitude, falling from $10^{14}$ at
the equator to $10^3$~cm$^{-3}$ at the upper boundary. At the
same time, the character of the temperature distribution is defined by the distance from the star. At $R \gtrsim 1$~AU, the temperature grows monotonically from the
equator to the disk atmosphere, while, at $R \lesssim 1$~AU,
the temperature first falls off, then grows with increasing $z$. This is related to the circumstance that, in the 
inner region of the disk, viscous dissipation begins to
provide a substantial input to the heating of the disk,
while, at $R \gtrsim 1$~AU, the dominant source of heating is
the UV radiation of the central star.

The panels in the middle row of Fig.~\ref{fig_2d} show the distributions of the density and temperature at $t = 3694$~yr,
corresponding to the time just before the enhancement of the accretion due to convection. Comparing
these distributions with the upper panels for $t = 0$, we
note that the number density (up to $10^{15}$~cm$^{-3}$) and the
temperature of the gas at the equator (up to $10^{3}$~K)
have grown substantially in the inner region of the
disk, $R<3$~AU. This is related to the accumulation of
matter in this region as a result of accretion from outer regions of the disk. The density and temperature distributions in the vicinity of $lg\, R(a.e.)=0.5$ become
more monotonic compared to these distributions at
the initial time.

The distributions in the lower panels of Fig.~\ref{fig_2d} are
for $t = 3889$~yr and correspond to the phase of developed convection and the transfer of matter from the
inner region. In the distribution of density in the vicinity of $\lg R = -0.3$, a region of enhanced density is
seen, corresponding to the convection propagation
front. The thickness of the disk inside this region is
slightly greater than in the stage of matter accumulation, due to the increased temperature in the convection zone. In all the distributions, the shaded areas
indicate the regions that satisfy the criterion of convective instability. The region of convective instability
is located near the equator and extends to the height $z/R \approx 0.03$.

\begin{figure}
\setcaptionmargin{5mm}
\onelinecaptionstrue
\includegraphics[angle=0,width=0.48\textwidth]{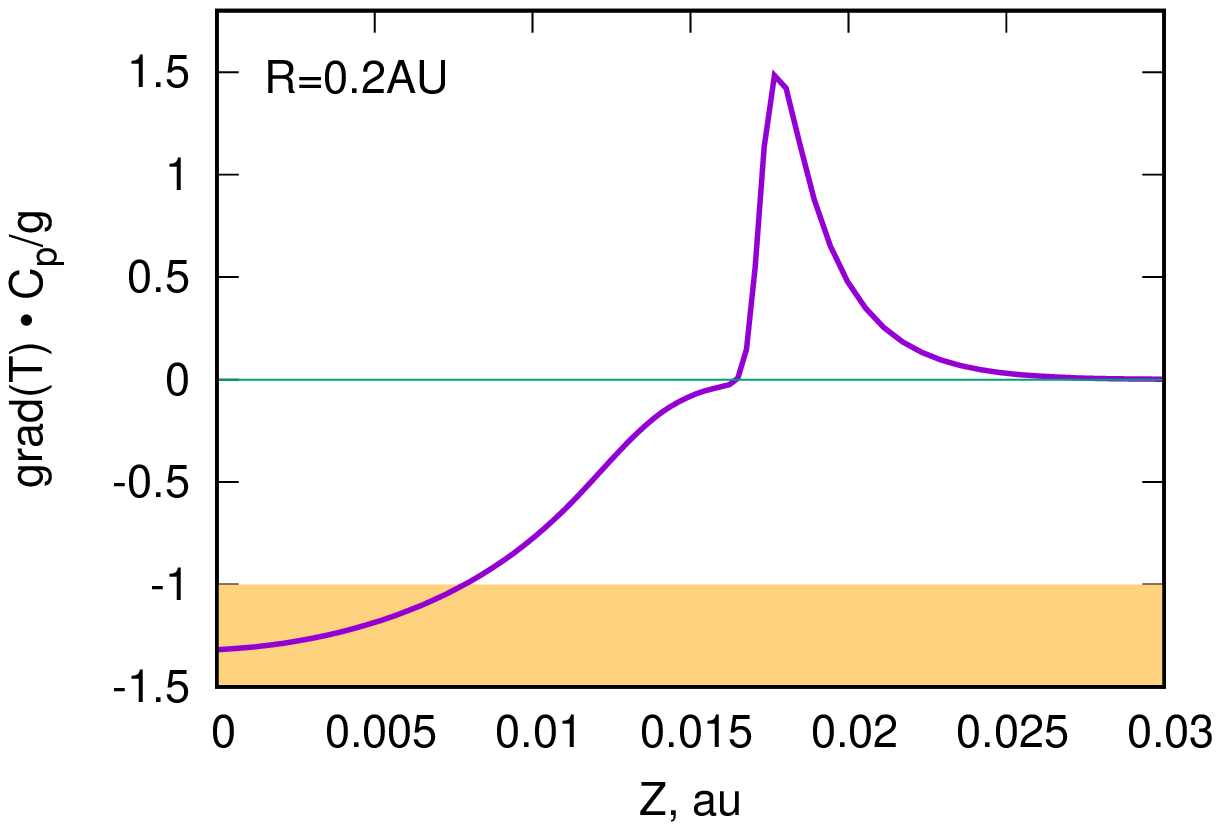}
\includegraphics[angle=0,width=0.48\textwidth]{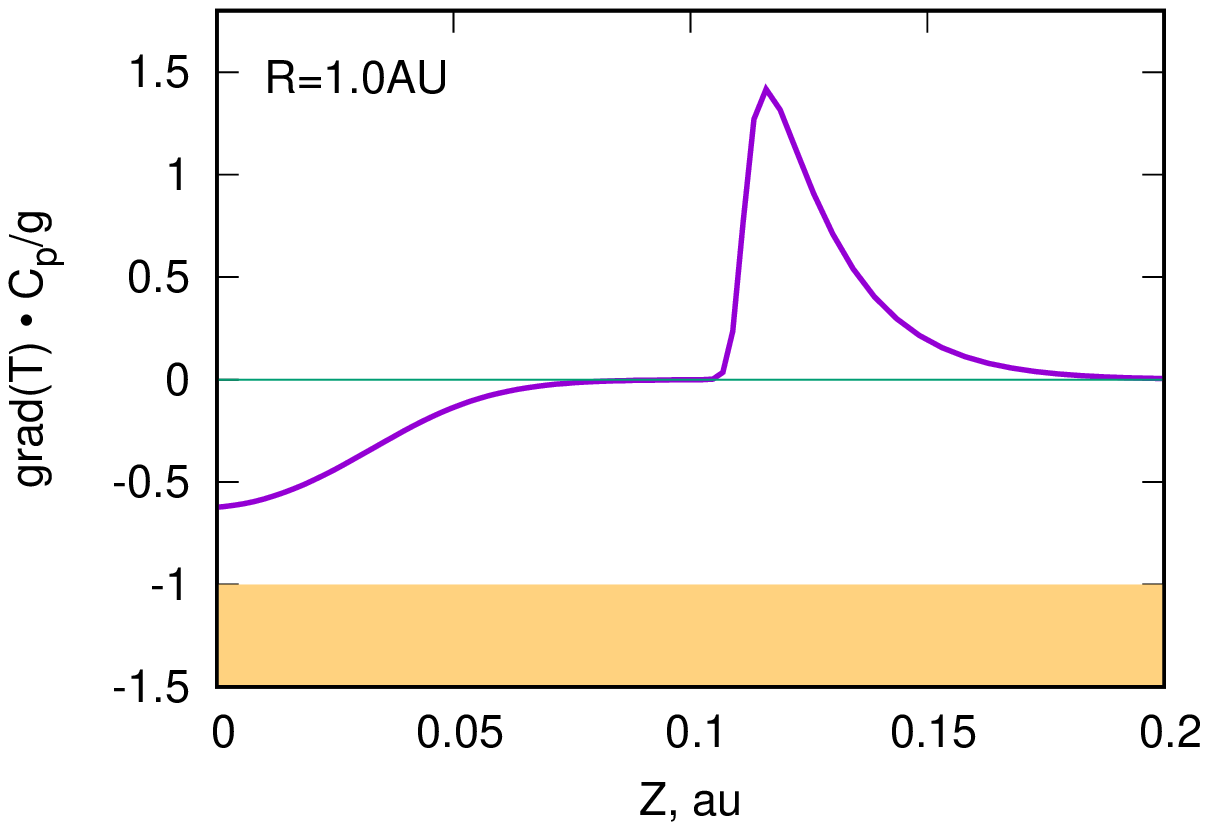}\\
\captionstyle{normal}
\caption{Ratio of the temperature gradient and adiabatic gradient as a function of the $z$ coordinate for $R=0.2$~AU (left) and
$R=1$~AU (right) for time 3889~yr, corresponding to the phase of an accretion burst. The orange strip corresponds to the region of
convective instability.}
\label{fig_vert}
\end{figure}

Figure~\ref{fig_vert} shows the distributions of the ratio of the
temperature gradient $dT/dz$ and the absolute value of
the adiabatic gradient $g(z)/c_p$ in the $z$ direction for two
radii at the stage of the burst in accretion ($t = 3889$~yr).
In the distribution for $R\approx0.2$~AU (the left panel of
Fig.~\ref{fig_vert}), there is a region where this ratio drops below $-1$, for which the condition of convective instability is
satisfied. At $R=1$~AU (the right panel of Fig.~\ref{fig_vert}), there
is also a negative temperature gradient in the vicinity
of the equator, but its absolute value does not exceed
the value of the adiabatic gradient. Therefore, this
region remains convectively stable at the given time.

\begin{figure}
\setcaptionmargin{5mm}
\onelinecaptionstrue
\includegraphics[angle=0,width=0.45\textwidth]{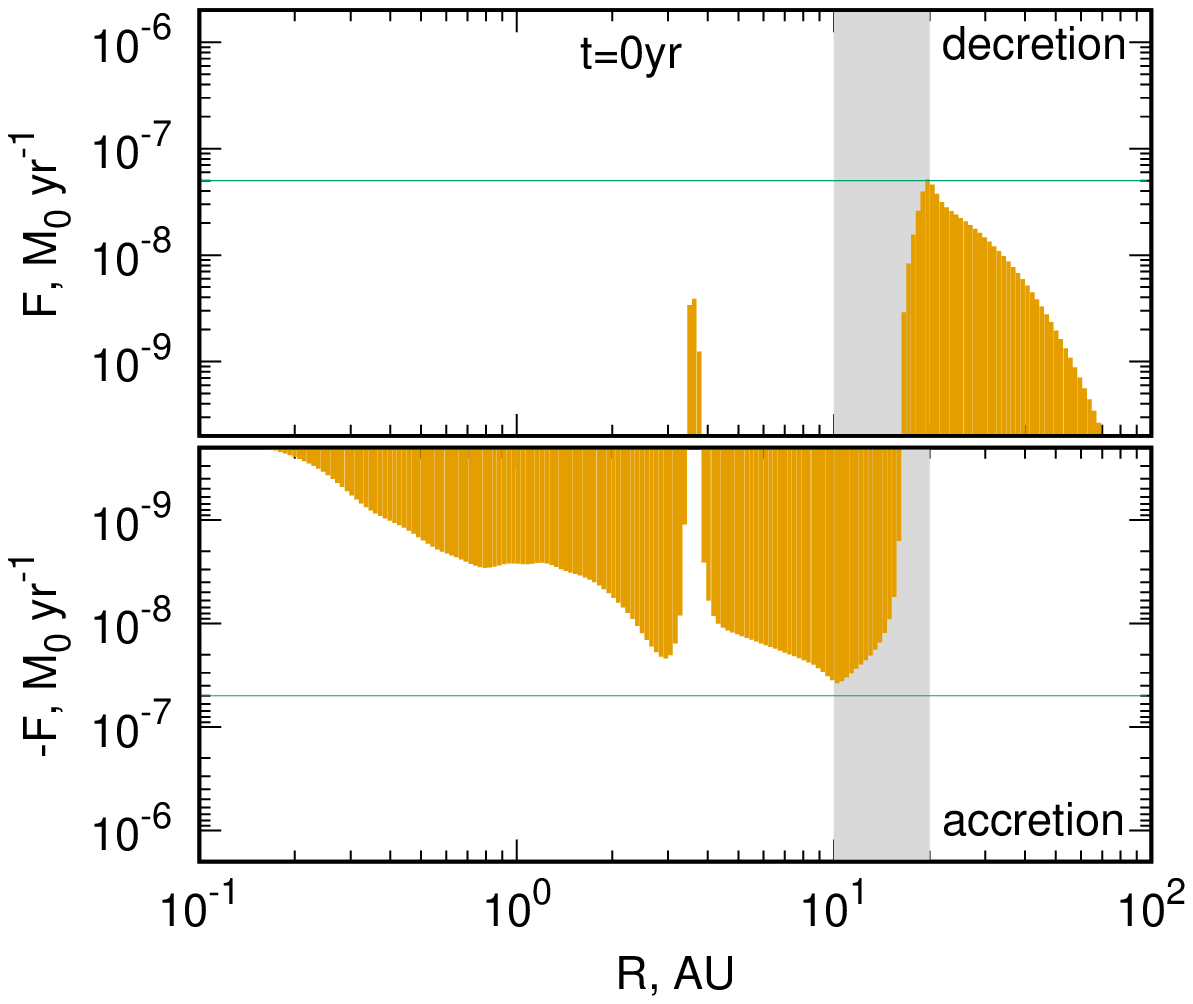}\\
\includegraphics[angle=0,width=0.45\textwidth]{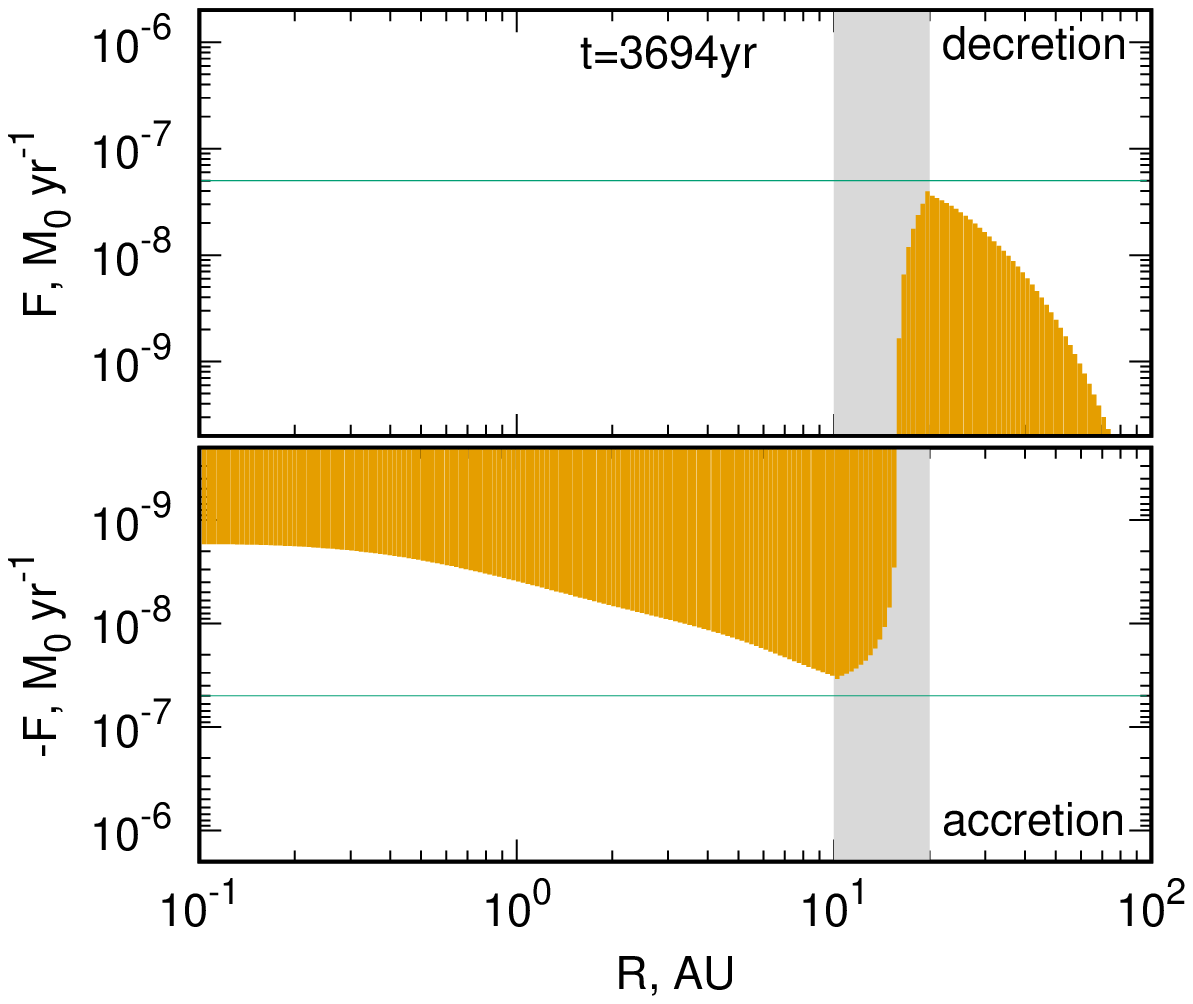}\\
\includegraphics[angle=0,width=0.45\textwidth]{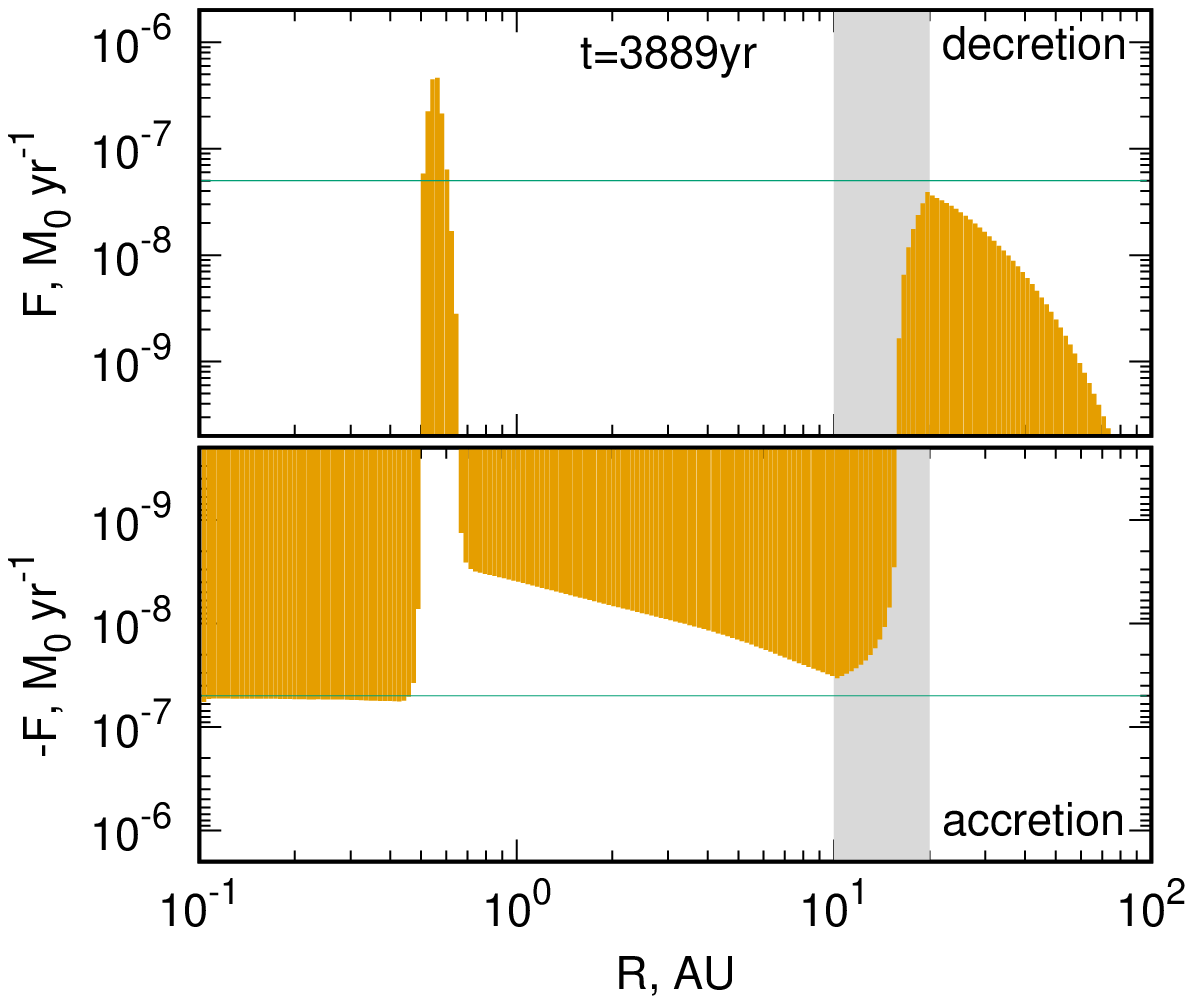}
\captionstyle{normal}
\caption{
Radial distributions of the accretion flux at time
zero (upper panel), just before the burst (middle panel),
and during the burst (lower panel). Positive fluxes (the
upper part of every distribution) correspond to flows from
the star, and negative fluxes (the lower part of the distribution) to flow toward the star. The time is measured from
the end of the previous accretion burst and is indicated at
the top of every panel. Horizontal lines correspond to
fluxes  $\pm 0.5\times 10^{-7}M_{\odot}$/yr. The vertical strip indicates the
region of accretion onto the star.}
\label{fig_flux}
\end{figure}

Figure~\ref{fig_flux} shows the dependences of the accretion
flux on the radius for the times just after the previous
accretion burst ($t=0$~yr), before the current burst
(3694~yr), and in the phase of matter transfer (3889~yr).

In the phase of matter accumulation ($0<t<3694$~yr), the disk can conveniently be divided into two
parts, corresponding to accretion and decretion (using
the terminology from~\cite{2004ARep...48..800T}). In the accretion part,
$R<15$~AU, the matter moves toward the star, while
the gas moves outward in the decretion part, $R>15$~AU. In our model, the border between these regions
is defined by the continuous influx of matter from the
envelope into the ring at $10$--$20$~AU. The absolute values of these fluxes on both sides of this border are close
to $0.5\times 10^{-7}M_{\odot}$/yr, i.e., to half the accretion rate from
the envelope. Thus, about half of the mass of the
accreting envelope enters the inner part of the disk,
while the other half goes into the outer part. Note,
however, that, in the model presented, the matter
accreted from the envelope is fed into the disk at the Keplerian speed, and with other assumptions, this picture will change. In general, this boundary can shift as the disk evolves. At time 3694 yr, the flow of matter in
the accretion part of the disk decreases toward the star,
taking its minimum value at the inner boundary of the
disk, thus providing evidence for the accumulation of
matter in this region.

In the phase of the accretion burst (lower panel of
Fig.~\ref{fig_flux}), a characteristic feature is seen in the flux distribution — a peak of positive flux with amplitude
$5\times 10^{-7}M_{\odot}$/yr near a radius of $0.5$~AU; this peak corresponds to the front of the convective region. As was
noted above, this front propagates outward. Inside a
radius of $0.5$~AU, the accretion flux exceeds the values
characteristic for the accumulation phase by more
than an order of magnitude. At the same time, the flux
distribution upstream of the convective front, i.e., at
$R > 0.5$~AU, remained the same.

\begin{figure}
\setcaptionmargin{5mm}
\onelinecaptionstrue
\includegraphics[width=0.32\textwidth]{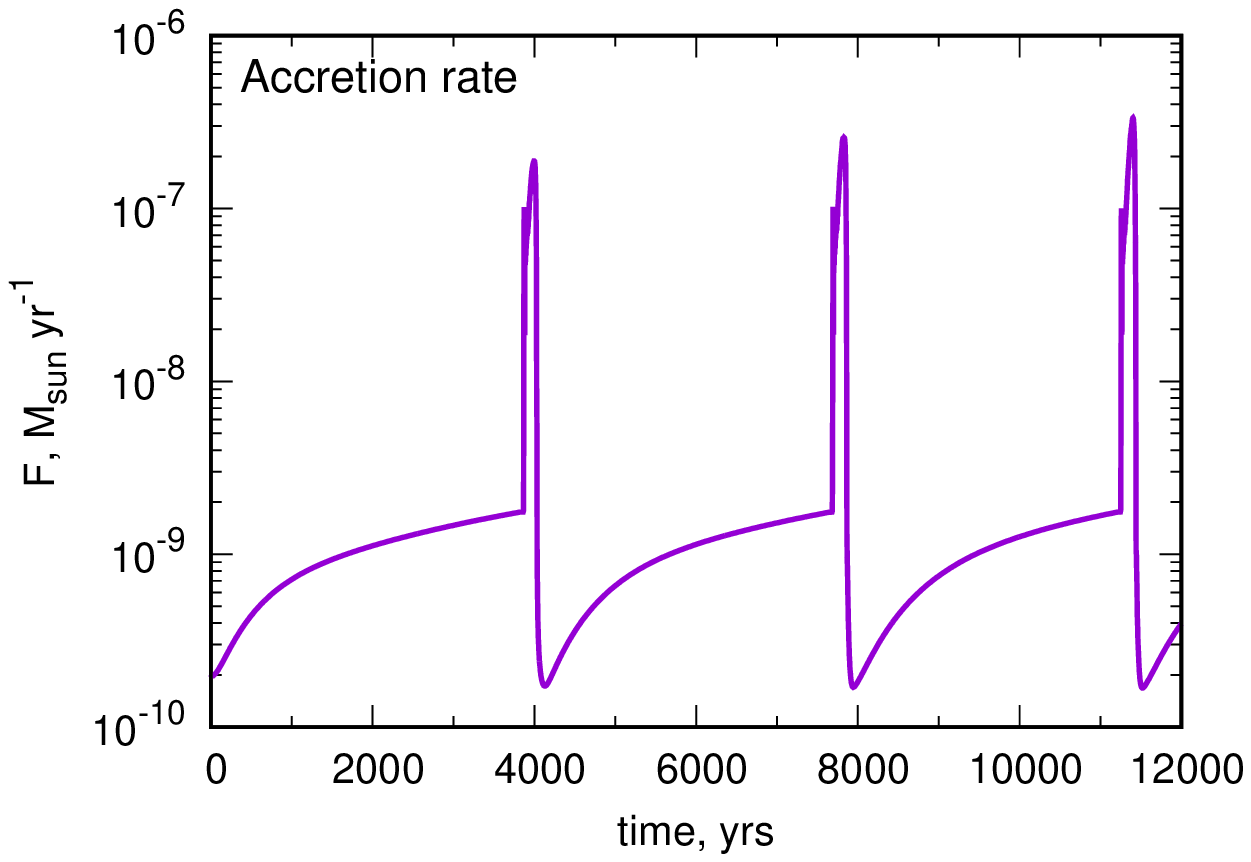}
\includegraphics[width=0.32\textwidth]{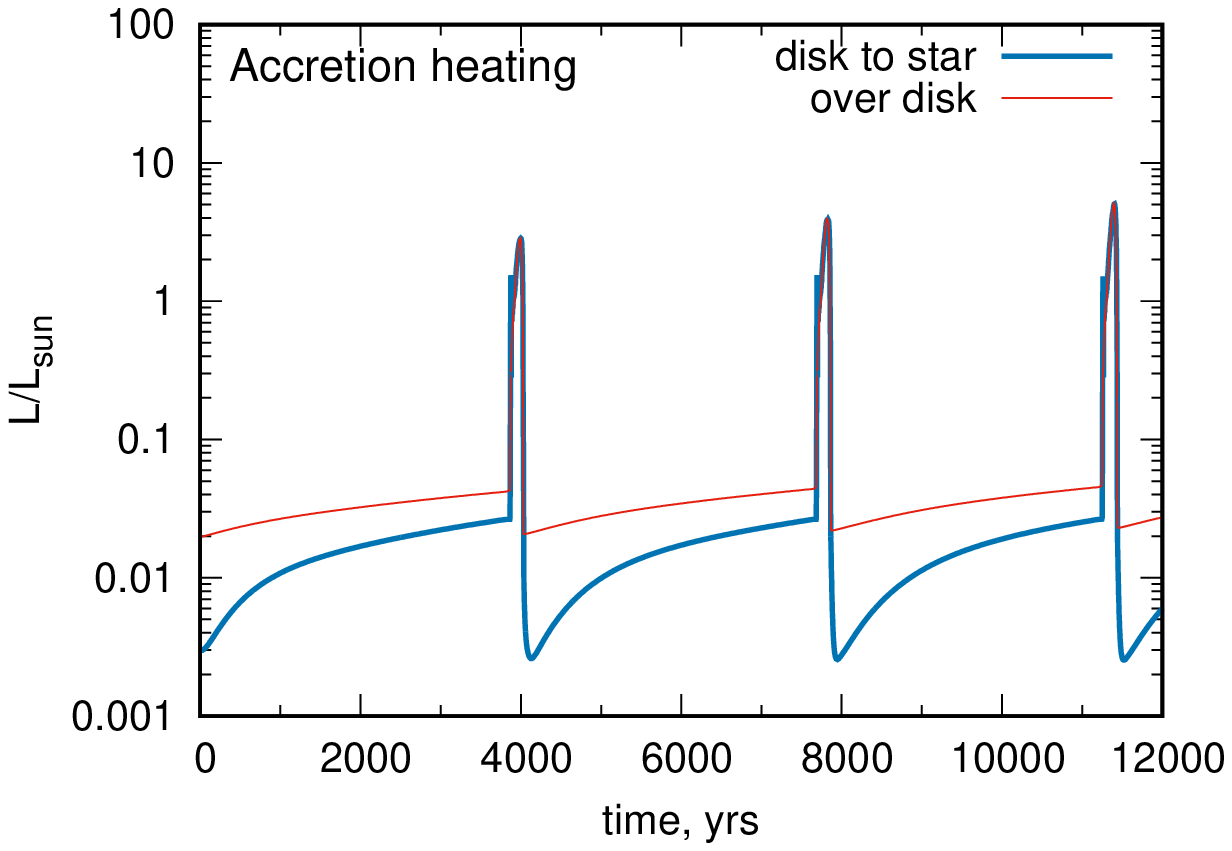}
\includegraphics[width=0.32\textwidth]{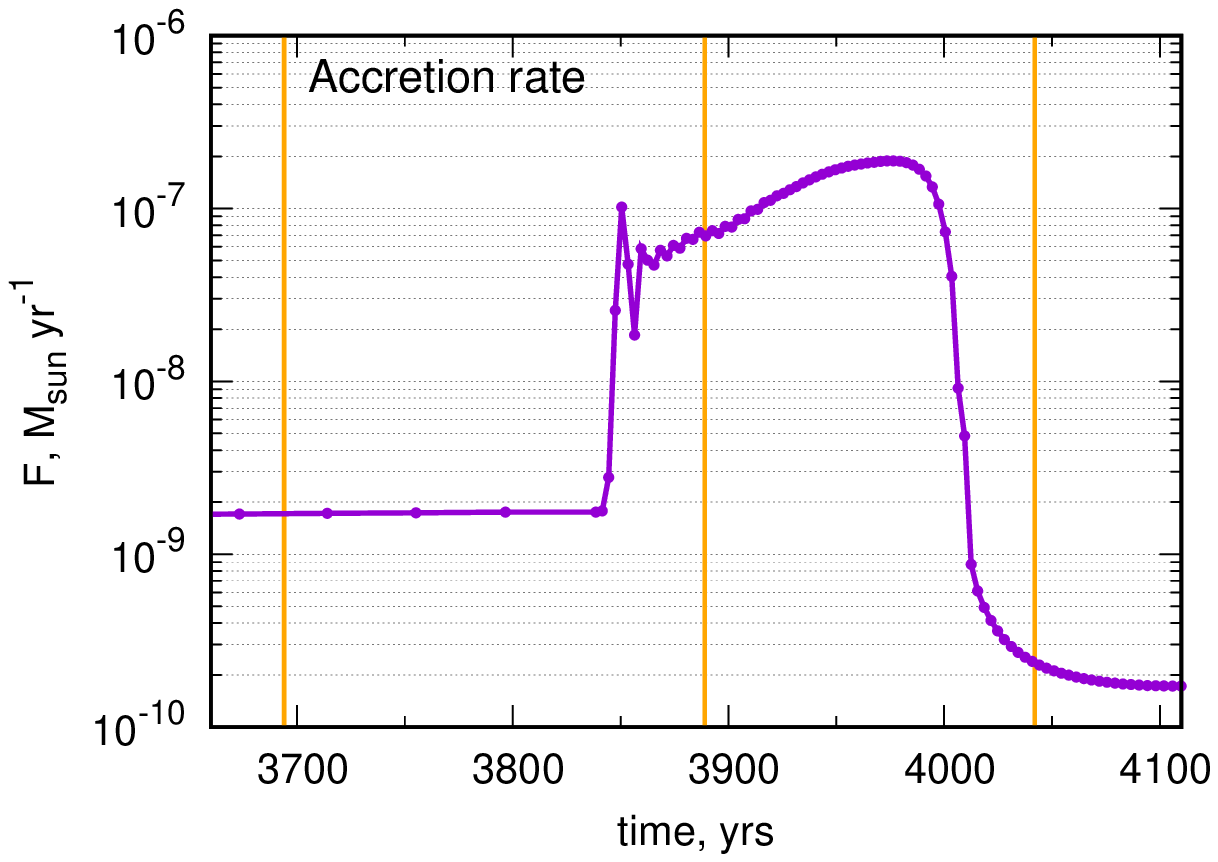}
\captionstyle{normal}
\caption{Dependence of the rate of accretion of the disk gas by the star (left panel) and the accretion luminosity (central panel)
for a time 12000 yr after the establishment of the episodic accretion regime. The rate of accretion onto the star during the burst phase
is shown in the right panel. The red curve in the middle panel shows the accretion luminosity of the entire disk, the blue curve
shows the luminosity associated with the accretion of gas from the inner boundary of the disk onto the star. The vertical (orange)
lines in the bottom panel correspond to times of 3694, 3889, and 4042 yr.}
\label{fig_curve}
\end{figure}

Figure~\ref{fig_curve} shows the rate of accretion onto the star
(left panel) and the accretion luminosity (middle
panel) for a time interval of 12000 yr after the establishment of the episodic accretion regime, as well as the
rate of accretion onto the star during the accretion burst phase (right panel). The accretion luminosity of
the entire disk (red curve in Fig.~\ref{fig_curve}) was computed by
integrating expression~\eqref{eq_vis} from the inner to the outer
boundary of the disk. The accretion luminosity associated with the accretion of gas from the inner boundary of the disk onto the star was computed using the
formula
\begin{equation}
L_{*}=\dfrac{1}{2}\dfrac{GM_{\odot}\dot{M}}{R_{\odot}},
\end{equation}
where $M_{\odot}$ and $R_{\odot}$ are the mass and radius of the star
(assumed to be the solar values). The dependences
given above show that the interval between accretion
outbursts is about 4000 yr, while the convective phase
itself lasts about 200 years. The rate of accretion onto
the star during the convective phase is approximately
two orders of magnitude higher than the rate of accretion in the quiescent period. The total rate of energy
release in the disk itself and at the surface of the star
during the active phase is about $6L_{\odot}$, 50--100 times
higher than the values characteristic for the phase of
gas accumulation in the disk.

\section{DISCUSSION}

Comparing the model rates of accretion and luminosity with those for young bursting FU Ori (FUor) and
EX Lup (EXor) objects~\cite{2014prpl.conf..387A}, we can conclude that our
model reproduces the characteristic duration of FUor
bursts (from several tens to hundreds of years) fairly
well. It is difficult to derive the duration of the quiescent phase between the bursts of FUors from observations; it varies from several thousand years~\cite{2018A&A...613A..18V} to several tens of thousands of years~\cite{2013MNRAS.430.2910S}. In our model, it is
close to the lower limit of observational estimates. On
the other hand, the maximum accretion rate during a
burst is close to the lower limit of observational estimates and is consistent only with data for NGC~722~\cite{2014prpl.conf..387A}. With regard to the maximum accretion rate and
accretion luminosity, our model is more consistent
with EXor-type bursts, but these objects have much
shorter and more frequent bursts lasting from several
months to several years~\cite{2014prpl.conf..387A}. Note that we have not yet
studied the entire range of possible model parameters.
In particular, we have not investigated the dependence
of the disk evolution on the law for gas accretion from
the envelope onto the disk. Test computations show
that with an increase in the rate of accretion from the
envelope, the time between bursts decreases, but the
duration and intensity of the bursts are preserved. Further investigation of the episodic accretion regime caused
by convective instability is needed for a more detailed
comparison with observations.

\begin{figure}
\setcaptionmargin{5mm}
\onelinecaptionstrue
\includegraphics[width=0.44\textwidth]{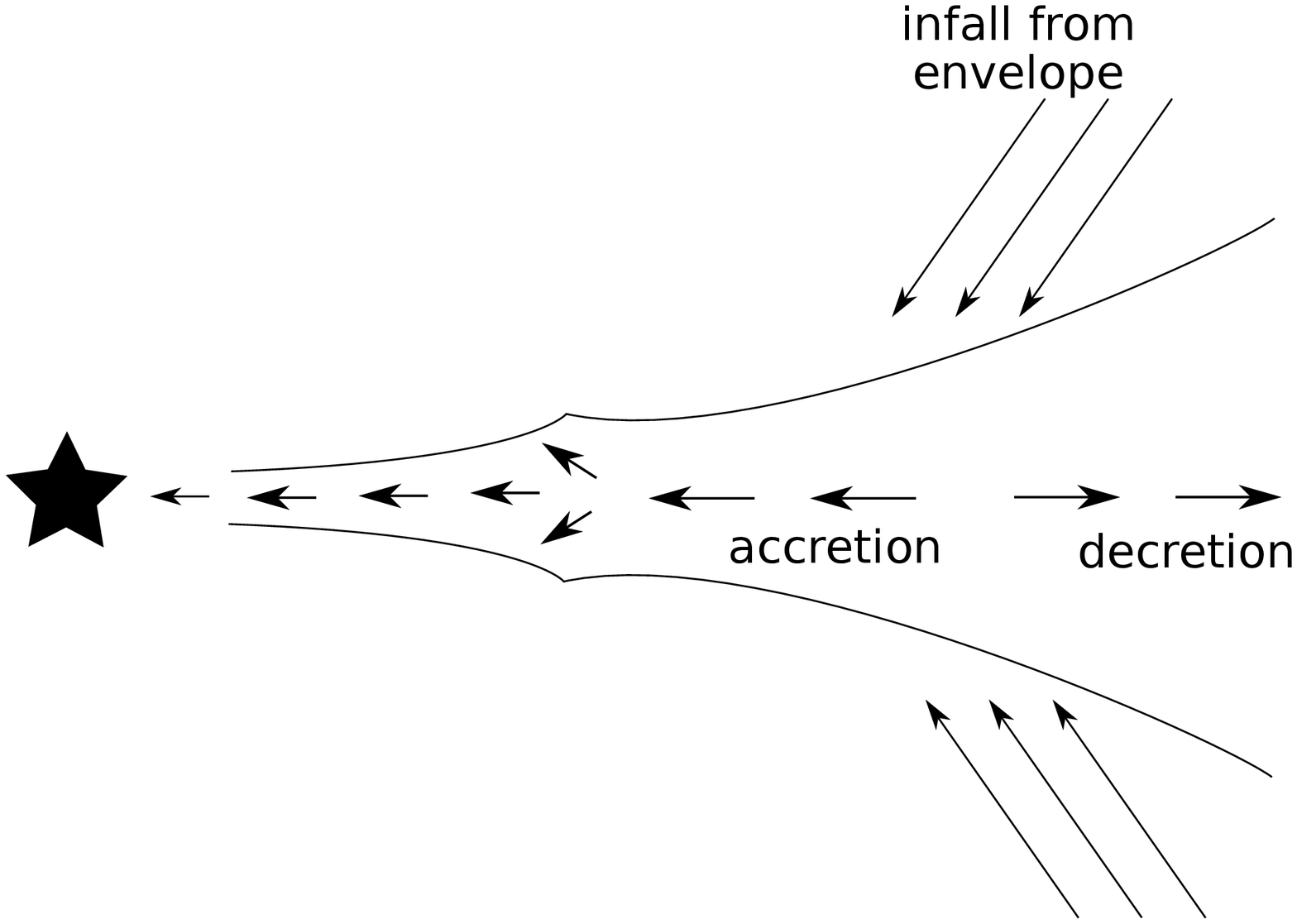}
\hfill
\includegraphics[width=0.44\textwidth]{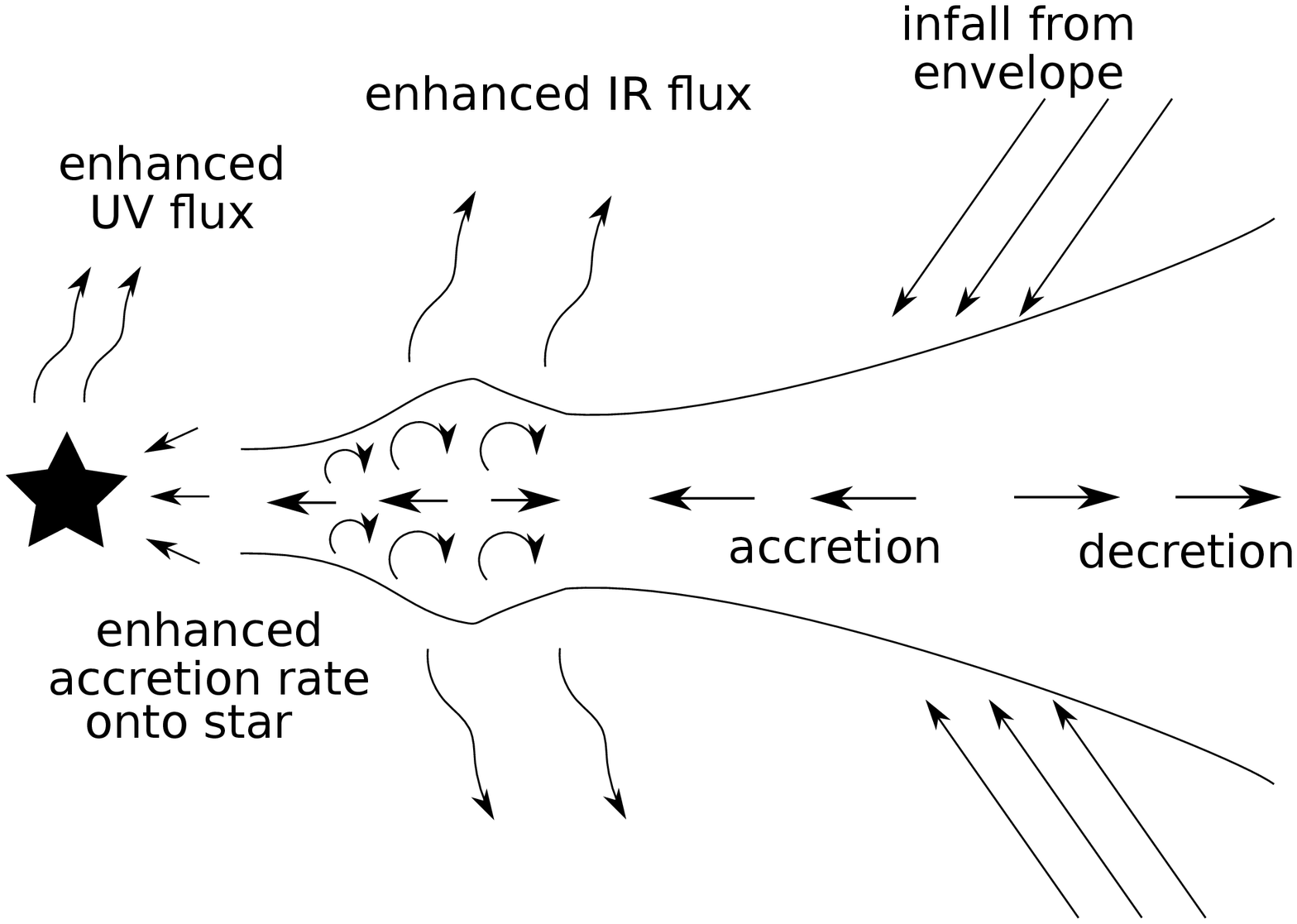}
\captionstyle{normal}
\caption{Sketch illustrating the burst-like nature of accretion in protoplanetary disks. The matter-accumulation phase is shown to
the left and the convective phase to the right.}
\label{fig_schemes}
\end{figure}

Note that our model is largely illustrative due to the
numerous physical approximations applied. Its main
purpose is to describe the evolution of the disk, and to
show the possible role of convection in enabling an episodic
accretion regime. This picture is presented schematically in Fig.~\ref{fig_schemes}.

We will now list a number of important issues that
underlie the model and require special attention in its
further development. We have assumed that, along
with convection, another unspecified mechanism acts
in the disk, which provides a "background" viscosity.
It is due to this mechanism that the transfer of mass
and angular momentum is realized in the quiescent
phase of the disk's evolution. Due to the background
viscosity, matter accumulates in the inner region of the
disk. In our model, convection is a trigger that facilitates the transfer of matter from the inner region of the
disk.

In our model, a fundamental condition for the
appearance of convectively unstable regions is an
increase in the opacity of the medium with increasing
temperature (see Fig.~\ref{figopac}). This is brought about by the
optical properties of the dust used in the model,
namely, the increase of the spectral absorption and
scattering coefficients with frequency. Our test computations show that, when using artificially specified
opacities that are independent of temperature, the
resulting temperature gradient is not strong enough to
ensure condition~\eqref{criterion}. The strong influence of the
opacity law on the appearance of convectively unstable
zones was noted in~\cite{1995AVest..29...99M,1996SoSyR..30..440M}, where the structures of
protoplanetary disks with various parameters were calculated and analyzed. It was shown, in particular, that
the evaporation of dust in the inner regions of the disk
reduces the opacity of the medium, eliminating the
appearance of convectively unstable regions. In our
model, dust evaporation is not taken into account, but
the maximum temperature can approach the characteristic dust evaporation temperatures. Therefore we
plan to take this process into consideration in the
future.

In our model, an increase in the opacity with
increasing temperature also leads to positive feedback
in the development of convection in the inner region
of the disk: an increase in the accretion rate leads to an
increased energy release, an increased temperature,
and, as a result, to more favorable conditions for the
appearance of convective instability. However, the
convectively unstable region is depleted relatively
quickly (in about hundred years). A strong decrease in
the density leads to a decrease in the accretion rate and
energy release, which ultimately leads to the restoration of convective stability in the inner region. Thus,
in this model, convection is self-sustaining only for
short periods of time in the inner regions of the disk,
and the background viscosity is important to ensure its
launch.

In our model, a prerequisite for the quasi-periodic
(burst) accretion regime is a steady inflow of matter
into the disk from the envelope. In our current realization of the model, the accretion of gas from the envelope was assumed to be constant, and the accretion
region was a ring between 10 and 20 AU. In reality, the
accretion rate and the position of the region of accretion of matter from the envelope are time dependent,
and this must be taken into account when constructing
models aimed at a more consistent comparison of the
model with observations. With a decreasing rate of
accretion from the envelope and depletion of the disk,
the inner regions become more transparent to their
own thermal radiation, and the rate of viscous heating
decreases. As a result of all this, the conditions for
convective instability cease to be satisfied.

An important element of the model is the method
used to compute the viscosity coefficient in the convectively unstable region. We used an approximation
in which the velocities of convective elements are calculated from the condition that all the thermal energy
released as a result of viscosity is transported by the
convective flow. In this approach, we do not allow for
the fact that some of the energy can be carried by radiation, i.e., the velocity of the convective elements is
somewhat overestimated. At the same time, when
reconstructing the disk structure in the polar direction, we did not take into account the convective
energy transfer and radiative energy transfer in the
radial direction; i.e., the temperature distribution was
found by taking into account radiative transfer in the
vertical direction only, in a steady-state approximation. In order to eliminate these inconsistencies, we
plan in the future to apply mixing-length theory,
which has successfully been used to take into account
convection in computations of stellar structure (see,
e.g., \cite{2004sipp.book.....H}).

The picture we have presented here is only one possible scenario for the episodic behavior of the accretion
around young stars. A wide variety of models have
already been proposed for FUors and EXors. For
example a two-dimensional hydrodynamical model is
presented in~\cite{2009ApJ...701..620Z}, in which the burst character of the
accretion is provided by magnetorotational instability,
while gravitational instability is responsible for the
influx of the matter into the inner regions of the disk
from its outer part. In~\cite{2006ApJ...650..956V}, the bursts of FUors are
explained by the infall of gravitationally bound fragments that form in the accretion disk and migrate
toward the star. We note also that, in order to explain
the observational manifestations of FUors and the
dynamics of the circumstellar gas, the presence of an
intense wind generated by the inner regions of the turbulent accretion disk is necessary~\cite{2019MNRAS.483.1663M}. In general,
recurrent accretion activity is a common occurrence in
accreting objects~\cite{Tutukov:2019}.

In conclusion, we note the need to reproduce the
obtained episodic character of accretion in the framework
of a hydrodynamical model. In a hydrodynamical
model, the development of convective zones can be
observed directly; i.e., there is no need for a phenomenological introduction of the viscosity coefficient in
the convective region. In a hydrodynamic model,
however, it is necessary to pay increased attention to
the computation of the thermal structure of the disk
and the selection of a suitable spatial grid, due to the
fact that the density gradients from the equator to the
disk atmosphere amount to many orders of magnitude.

\section*{ACKNOWLEDGMENTS}
We thank the referee for valuable comments and constructive suggestions for improvement of the paper. We also
thank A.~B.~Makalkin for useful discussions.

\section*{FUNDING}

This project was supported by the Russian
Foundation for Basic Research (project 17-02-00644).


%
%

\selectlanguage{english}
\bibliography{maiksamp}

\it{Translated by L. Yungelson}
\newpage
%
%


\end{document}